\documentclass[sigconf]{acmart}

\usepackage{array,graphicx,epsfig,amssymb,amsfonts}
\usepackage{bbm,amsmath,epstopdf,algorithm,subfigure, multirow}
\usepackage[noend]{algpseudocode}
\usepackage{xcolor}
\usepackage{float}
\usepackage{dblfloatfix}
\usepackage{url}
\usepackage{pifont}
\usepackage[framemethod=1]{mdframed}
\allowdisplaybreaks

\newcommand{\be}{\begin{equation}}
	\newcommand{\ee}{\end{equation}}
\newcommand{\bee}{\begin{eqnarray}}
	\newcommand{\eee}{\end{eqnarray}}
\newcommand{\bse}{\begin{subequations}}
	\newcommand{\ese}{\end{subequations}}
\newcommand{\nnb}{\nonumber}

\newcommand{\cmark}{\ding{51}}
\newcommand{\xmark}{\ding{55}}

\newcommand{\specialcell}[2][c]{%
	\begin{tabular}[#1]{@{}c@{}}#2\end{tabular}}

\setcopyright{rightsretained}

\ifodd1
    
    \newcommand{\com}[1]{\textbf{\color{blue} (COMMENT: #1)}} 
\else
    
    \newcommand{\com}[1]{}
\fi

\begin{document}
\title{Minimizing Age-of-Information with Throughput Requirements\\in Multi-Path Network Communication}
\author{Qingyu Liu, Haibo Zeng}
\affiliation{
	\department{Electrical and Computer Engineering}
	\institution{Virginia Tech}
	}
\author{Minghua Chen}
\affiliation{
	\department{Information Engineering}
	\institution{The Chinese University of Hong Kong}
	}	

\ccsdesc[500]{Mathematics of computing~Network flows}
\ccsdesc[500]{Networks~Network resources allocation}

\keywords{Age-of-information, multi-path routing, time-critical network flow}

\begin{abstract}
We consider the scenario where a sender periodically sends a batch of data to a receiver over a multi-hop network, possibly using multiple paths. Our objective is to minimize peak/average Age-of-Information (\textsf{AoI}) subject to throughput requirements. The consideration of batch generation and multi-path communication differentiates our \textsf{AoI} study from existing ones. We first show that our \textsf{AoI} minimization problems are NP-hard, but only in the weak sense, as we develop an optimal algorithm with a pseudo-polynomial time complexity. We then prove that minimizing \textsf{AoI} and minimizing maximum delay are ``roughly" equivalent, in the sense that any optimal solution of the latter is an approximate solution of the former with bounded optimality loss. We leverage this understanding to design a general approximation framework for our problems. It can build upon any $\alpha$-approximation algorithm of the maximum delay minimization problem, e.g., the algorithm in~\cite{misra2009polynomial} with $\alpha=1+\epsilon$ given any user-defined $\epsilon>0$, to construct an $(\alpha+\mathsf{c})$-approximate solution for minimizing \textsf{AoI}. Here $\mathsf{c}$ is a constant depending on the throughput requirements. Simulations over various network topologies validate the effectiveness of our approach.  
\end{abstract}

\copyrightyear{2019} 
\acmYear{2019} 
\setcopyright{acmcopyright}
\acmConference[Mobihoc '19]{The Twentieth ACM International Symposium on Mobile Ad Hoc Networking and Computing}{July 2--5, 2019}{Catania, Italy}
\acmBooktitle{The Twentieth ACM International Symposium on Mobile Ad Hoc Networking and Computing (Mobihoc '19), July 2--5, 2019, Catania, Italy}
\acmPrice{15.00}
\acmDOI{10.1145/3323679.3326502}
\acmISBN{978-1-4503-6764-6/19/07}

\maketitle

\section{Introduction}\label{sec:introduction}
Age-of-Information (\textsf{AoI}) is a critical networking performance metric for periodic services that require timely transmissions. Kaul \emph{et al.}~\cite{kaul2011minimizing} measures the \textsf{AoI} as the time that elapsed since the last received update was generated. Upon receiving a new packet with updating information, the AoI drops to the elapsed time since the packet generation; otherwise, it grows linearly in time. In this paper, we study fundamental \textsf{AoI}-minimization problems of supporting a periodic transmission task over a multi-hop network. The task requires a sender to send a batch of data (packets) periodically to a receiver, possibly using multiple paths. Our objective is to minimize peak/average \textsf{AoI} subject to both a minimum and a maximum throughput requirement, by jointly optimizing throughput and multi-path routing strategy. We assume the amount of data in the batch is fixed, hence the throughput (the ratio of the volume of the data batch over the task activation period) only varies with the task activation period. 

\begin{table*}[]
	\centering
	\caption{Compare our \textsf{AoI} study with existing ones.}
	\vspace{-2ex}
	\label{tab:relatedwork}
	\scalebox{0.9}{\begin{tabular}{|c|c||c|c|c|c|c|c|c|c|}
			\hline
			\multicolumn{2}{|l||}{} & \cite{kaul2011minimizing,kaul2012real,sun2017update} & \cite{sun2018age} & \cite{huang2015optimizing} & \cite{talak2018can} & \cite{kadota2018optimizing} & \cite{talak2017minimizing,talak2018ISIT} & \cite{talak2018Mobihoc} & Our work\\ \hline\hline
			\multirow{2}{*}{\specialcell{Objective of\\Optimization}} & Minimize peak \textsf{AoI} & \xmark & \cmark & \cmark & \cmark & \xmark & \cmark & \cmark & \cmark \\ \cline{2-10}
			& Minimize average \textsf{AoI} & \cmark & \cmark & \xmark & \cmark & \cmark & \cmark & \cmark & \cmark \\ \hline \hline
			\multirow{4}{*}{\specialcell{Design Space of\\Optimization}} & \specialcell{Multi-path routing strategy} & \xmark & \xmark & \xmark & \xmark & \xmark & \xmark & \xmark & \cmark \\ \cline{2-10}
			& \specialcell{Information generation rate$^*$} & \cmark & \xmark & \cmark & \cmark & \cmark & \xmark & \cmark & \cmark \\ \cline{2-10}
			& \specialcell{Link scheduling policy} & \xmark & \xmark & \xmark & \xmark & \cmark & \cmark & \cmark & \xmark \\ \cline{2-10}
			& \specialcell{Queuing disciplines} & \xmark & \cmark & \cmark & \cmark & \xmark & \xmark & \xmark & \xmark \\ \hline\hline
			Other Results & Compare \textsf{AoI} with delay & \cmark & \xmark & \cmark & \cmark & \xmark & \xmark & \xmark & \cmark \\ \hline
		\end{tabular}}
		\linebreak
		\footnotesize \emph{Note}. $^*$: Under our system model, the information generation rate is equivalent to the achieved throughput.
		\vspace{-1ex}
\end{table*}

\noindent
\textbf{Motivations}. Our study is motivated by leveraging a network platform with limited resources to support periodic transmission tasks that are sensitive both to throughputs and to end-to-end delays. A  particular example is offloading real-time image-processing tasks in edge computing, with AoI taken into consideration.

Nowadays the blending of mobile/embedded devices and image processing is taking place, where deep learning is often involved to make devices smarter. Since deep learning is resource-heavy, while the mobile/embedded device is resource-constrained, in general those tasks cannot be executed locally on mobile/embedded devices timely as well as frequently. 
The widely-adopted solution is to leverage nearby powerful edge servers for workload offloading. For example, Ran \emph{et al.}~\cite{ran2017delivering} develop an Android application of real-time object detection. If running locally on the phone for 30 minutes, it processes images at a 5 FPS rate and consumes $25\%$ battery. As a comparison, if running remotely on a server, it processes images at the rate of 9 FPS and consumes $15\%$ battery.

From~\cite{ran2017delivering} we note that the majority (over $95\%$) of the total delay of running tasks remotely is the networking delay. Therefore, to offload the resource-heavy image-processing tasks to an edge computing platform for processing in real-time, time-critical offloading algorithms are vital to efficiently and timely utilize available resources. As the results of the offloaded tasks need to be sent to control units, e.g., at the end users or the edge computing nodes, for real-time actions, e.g., cyber-phystical system control, it is important to minimize the age of the information.

We compare our \textsf{AoI} study with existing ones in Tab.~\ref{tab:relatedwork}. Details refer to Sec.~\ref{sec:related}. In summary, the consideration of batch generation and multi-path communication differentiates our \textsf{AoI} study from existing ones. We study multi-path network communication problems of minimizing peak/average \textsf{AoI} for periodically transmitting a batch of data, subject to both a minimum and a maximum throughput requirement. We claim the following \textbf{contributions}.

$\mbox{\ensuremath{\rhd}}$ Comparing minimizing peak/average \textsf{AoI} with minimizing maximum delay: (i) 
we show that the optimal solution of the former can achieve a throughput that is different from, but always no smaller than, that achieved by the optimal solution of the latter (Lem.~\ref{lem:delay-aoi}). This result is consistent with our observation that \textsf{AoI} is a metric simultaneously considering maximum delay and throughput; (ii) we show that the optimal solution of the latter can be suboptimal to the former (Lem.~\ref{lem:delay-aoi}), but with a bounded optimality loss (Lem.~\ref{lem:delay-aoi-gap}).

$\mbox{\ensuremath{\rhd}}$ Comparing minimizing peak \textsf{AoI} with minimizing average \textsf{AoI}: (i) we prove that the optimal solution of the former can be suboptimal to the latter, and vice versa, but both with bounded optimality losses (Lem.~\ref{lem:peak-average}); (ii) we show that the optimal solution of the former can achieve a throughput (resp. maximum delay) that is different from, but always no smaller than, that achieved by the optimal solution of the latter (Lem.~\ref{lem:peak-average}). Thus, the problem of minimizing peak \textsf{AoI} may carry more flavor on throughput and less on maximum delay, compared to that of minimizing average \textsf{AoI}.

$\mbox{\ensuremath{\rhd}}$ We observe that both minimizing peak \textsf{AoI} and minimizing average \textsf{AoI} are challenging, because (i) we prove that both minimal peak \textsf{AoI} and minimal average \textsf{AoI} are non-monotonic, non-convex, and non-concave with throughput theoretically (Lem.~\ref{thm:age-T}), and (ii) we prove that both problems are NP-hard (Lem.~\ref{lem:np-hard}), but in the weak sense (Thm.~\ref{thm:weak-NP-hard}), as we design an algorithm to solve them optimally in a pseudo-polynomial time (Sec.~\ref{subsec:pseudo-algorithm}). 

$\mbox{\ensuremath{\rhd}}$ We further leverage our understanding on comparing \textsf{AoI} with maximum delay to develop an approximation framework (Thm.~\ref{thm:MLP-AOI}). It can build upon any $\alpha$-approximation algorithm of the maximum delay minimization problem, e.g., the algorithm in~\cite{misra2009polynomial} with $\alpha=1+\epsilon$ given any user-defined $\epsilon>0$, to construct an $(\alpha+\mathsf{c})$-approximate solution for minimizing \textsf{AoI}. Here $\mathsf{c}$ is a constant depending on the throughput requirements. Our framework has the same time complexity as that of the used  $\alpha$-approximation algorithm, and suggests a new avenue for designing approximation algorithms for minimizing \textsf{AoI} in the field of multi-path network communication.

$\mbox{\ensuremath{\rhd}}$ We conduct extensive simulations to evaluate our proposed approaches (Sec.~\ref{sec:experiments}). Empirically (i) our optimal algorithm obtains more than $3\%$ \textsf{AoI} reduction compared to our approximation framework, if the range of task activation period increases by $1$. However, (ii) our approximation framework has a constant running time of $0.06$s, while the running time of our optimal algorithm can increase by $0.12$s if the range of task activation period increases by $1$.

\section{Related Work}\label{sec:related}
\begin{figure}[!t]
	\centering
	\vspace{-2ex}
	\includegraphics[width=1.1\columnwidth]{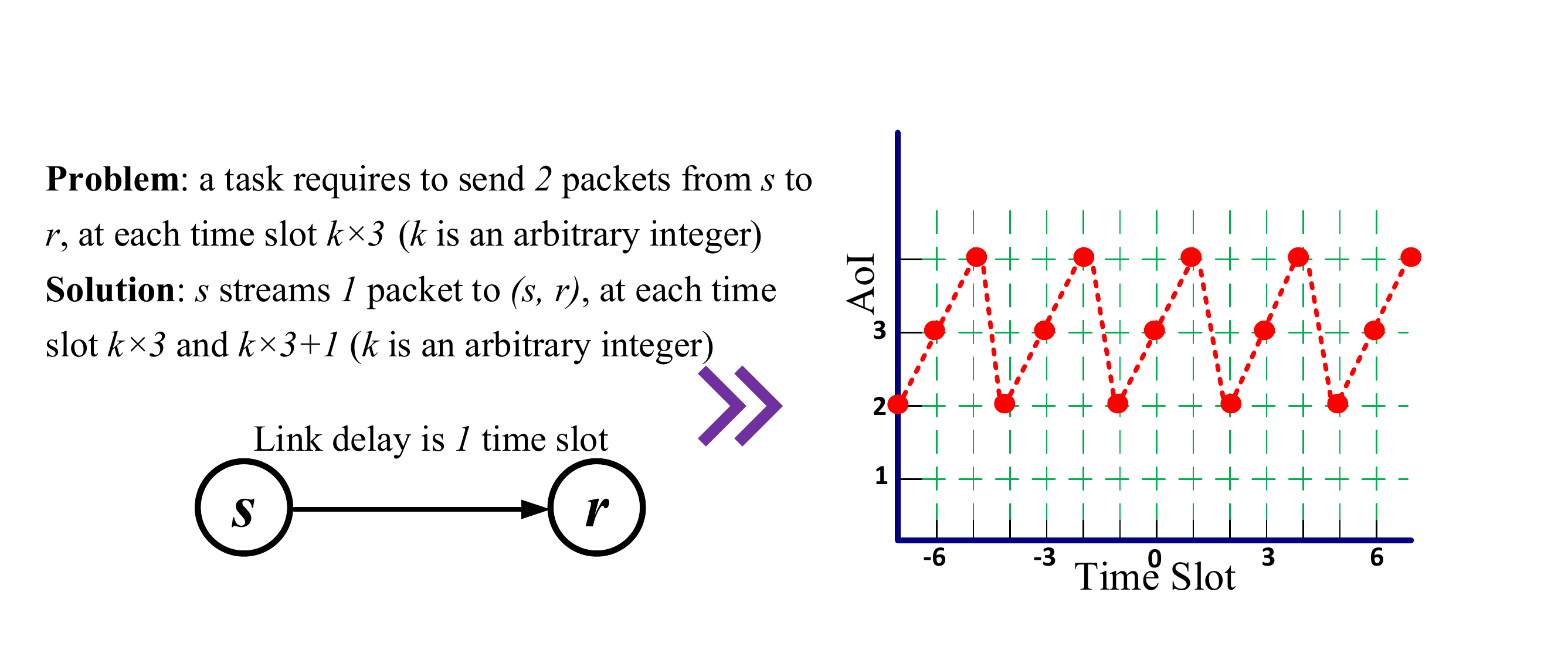}
	\vspace{-6ex}
	\caption{An illustrative example of our batch-based \textsf{AoI}. Sender $s$ generates a batch of two packets at each slot $3k,\forall k\in\mathbb{Z}$. It sends the two packets one-by-one over link $(s,r)$ to the receiver $r$; the link transmission incurs one-slot delay. Our batch-based \textsf{AoI} drops only when all packets from the same batch are received by $r$. Hence, as receiver $r$ receives all the two packets in a batch at each slot $3k+2,\forall k\in\mathbb{Z}$, the batch-based \textsf{AoI} at each slot $3k+2$ drops to $2$, i.e., the elapsed time since the generation of the last received batch. The batch-based \textsf{AoI} at all the other slot grows linearly.}	
	\label{fig:example}
\end{figure}

Since introduced by~\cite{kaul2011minimizing}, \textsf{AoI} has been studied theoretically and experimentally by various studies, which are summarized in Tab.~\ref{tab:relatedwork}. We differ from existing \textsf{AoI} studies in two aspects, i.e., the problem design space and the \textsf{AoI} definition. 

We note that multi-path routing is a basic paradigm of network communication. It is a natural extension of the single-path routing when streaming a high volume of traffic while avoid link traffic congestions. Many existing studies, e.g.,~\cite{wang2005cross,wang2011cost,wang2017sending}, have shown that multi-path routing can provide better \textsf{QoS}, e.g., larger throughput, than the single-path routing. To our best knowledge, this is the first work to optimize the multi-path routing strategy to minimize \textsf{AoI}.

Besides, existing studies define \textsf{AoI} at the packet level. Such definitions assume that \textsf{AoI} can be updated by receiving any packet, which are reasonable in status update systems. However, in our task-level study, we fairly assume that the receiver can reconstruct information of one task period and hence update \textsf{AoI} accordingly, only after it receives all the packets in a batch belonging to that task period. By this assumption, our batch-based \textsf{AoI} drop only upon successful reception of the complete batch of data belonging to one task period, and increase linearly otherwise. We give an illustrative example in Fig.~\ref{fig:example}, assuming slotted data transmissions.

Note that our problems are challenging, further compared to existing time-critical multi-path communication studies. Our problems minimize \textsf{AoI} with throughput requirements. Thus the task activation period (the ratio of the amount of data in the batch over throughput) is a decision variable under our setting. In contrast, to our best knowledge, existing time-critical multi-path communication problems minimize maximum delay given a fixed task activation period. Such problems include the quickest flow problem and the min-max-delay flow problem. Here the maximum delay is the time of sending a complete batch of data from the sender to the receiver, and clearly that \textsf{AoI} is a metric jointly considering maximum delay and task activation period. 

\emph{Quickest flow problem}~\cite{lin2015quickest,saho2017cancel}. Given an amount of data, it finds the minimum time needed to send them from a sender to a receiver, and the corresponding multi-path routing solution. This problem assumes that the task activation period is infinitely large, and is polynomial-time solvable under our setting~\cite{lin2015quickest}.

\emph{Min-max-delay flow problem}~\cite{misra2009polynomial,liu2018tale}. Given a sender-receiver pair and an amount of data, it finds a set of sender-to-receiver paths such that the maximum path delay of the set of paths is minimized while the aggregate bandwidth of the set of paths is no smaller than the given amount of data. This problem (also known as problem \textsf{OMPBD} studied in~\cite{misra2009polynomial}) assumes that the task activation period is one unit of time, and is NP-hard under our setting~\cite{misra2009polynomial}.

If $D$ is the given amount of data, note again that for the task activation period $T$, under our setting, we have $T=+\infty$ for the quickest flow problem, $T=1$ for the min-max-delay flow problem, but $D/R_u\le T\le D/R_l$ for our problems where $R_u$ (resp. $R_l$) is our maximum (resp. minimum) throughput requirement. Thus for the throughput $R$, we have $R\to 0$ for the quickest flow problem, $R=D$ for the min-max-delay flow problem, but $0< R_{l}\le R\le R_{u}\le D$ for our problems. According to Lem.~\ref{lem:age-delay} introduced later, given $R_{l}=R_{u}$, minimizing \textsf{AoI} is equivalent to minimizing maximum delay. This implies that the quickest flow problem ($R_l=R_u\to 0$) and the min-max-delay flow problem ($R_l=R_u=D$) are special cases of our problems. However, although exact algorithms for the quickest flow problem~\cite{lin2015quickest} and approximation algorithms for the min-max-delay flow problem~\cite{misra2009polynomial} have been developed, it is still not clear how to solve our problems even given $0<R_{l}=R_{u}\le D$. Moreover, according to Lem.~\ref{lem:delay-aoi}, given $0<R_{l}<R_{u}\le D$, minimizing \textsf{AoI} can differ from minimizing maximum delay. Overall, we observe that our \textsf{AoI}-minimization problems are uniquely challenging.

In the literature there exist some other time-critical periodic communication studies. For example, Hou et al.~\cite{hou2009theory} propose scheduling policies for a set of senders to be feasible with respect to the delay constraint, throughput constraint, and wireless channel reliability constraint.
Deng et al.~\cite{deng2017timely} further conduct a complete study on the similar timely wireless flow problem but assuming a more general traffic pattern. Those studies~\cite{hou2009theory,deng2017timely} are of little relevance with our problems, because their focus is the wireless link scheduling policy optimization. Differently, we focus on the throughput optimization and the multi-path routing strategy optimization.
\section{System Model}\label{sec:definition}
\begin{table}[]
	\centering
	\caption{Summary of important notations.}
	\vspace{-2ex}
	\label{tab:notations}
	\scalebox{0.9}{\begin{tabular}{|c|c|}
		\hline
		$\mathcal{T}(f)$ & \specialcell{Task activation period of a solution $f$}\\\hline
		\specialcell{$\Lambda_p(f)$ (resp.\\$\Lambda_a(f)$, $\mathcal{M}(f)$)} & \specialcell{Peak \textsf{AoI} (resp. Average \textsf{AoI},\\Maximum delay) of $f$}\\
		\hline
		$R_l$ (resp. $R_u$) & \specialcell{Input minimum (resp. maximum)\\throughput requirement}\\
		\hline
		\specialcell{$\Lambda_p^R$ (resp. $\Lambda_a^R$,\\$\mathcal{M}^R$)} & \specialcell{Minimal peak \textsf{AoI} (resp. Minimal average\\\textsf{AoI}, Minimal maximum delay) that can\\be achieved by any feasible periodically\\repeated solution with a throughput of $R$}\\
		\hline
		\specialcell{$R_p$ (resp. $R_a$,\\$R_m$)} & \specialcell{Achieved throughput that is optimal to our\\peak \textsf{AoI} (resp. average \textsf{AoI}, maximum\\delay) minimization problem}\\
		\hline
	\end{tabular}}
	\vspace{-2ex}
\end{table}
\subsection{Preliminary}
We consider a multi-hop network modeled as a directed graph $G \triangleq (V,E)$ with $|V|$ nodes
and $|E|$ links. We assume slotted data transmissions. Each link $e \in E$ has a bandwidth $b_e$ and a delay $d_e$. At the beginning of each time slot, each link $e$ can stream an amount of data that is no larger than the bandwidth $b_e\in\mathbb{R},b_e\ge 0$ ($b_e$ is a non-negative real number) to it, and this data experiences a delay of $d_e\in\mathbb{Z}^+$ ($d_e$ is a positive integer) slots to pass it. Besides, we assume that each node $v\in V$ can hold an arbitrary amount of data at each time slot. For easier reference, in this paper, we use ``at time $t$" to refer to ``at the beginning of the time slot $t$". We focus on a task that requires a periodic data transmission. Specifically, given that the \textbf{task activation period} is $T\in\mathbb{Z}^+$, the task will generate $D\in\mathbb{R},D>0$ amount of data at a sender node $s\in V$ at time $kT$ for each $k\in\mathbb{Z}$ ($k$ is an integer), and is required to transmit them to a receiver node $r\in V\backslash\{s\}$, possibly using multiple paths. Because we assume no data loss during transmission, the throughput incurred by a task activation period of $T$ is $D/T$.  

We aim to obtain a ``fresh" multi-path routing solution that is \emph{periodically repeated} to periodically send the batch of data. Here the ``freshness" is evaluated by \textsf{AoI} that is a function of the end-to-end networking delay (see our formula~\eqref{eqn:age}). It is well-known that the networking delay is mainly composed of propagation delay, transmission delay, and queuing delay. Similar to the discussions in~\cite{bai2012dear}, we remark that the slotted data transmission model can take all different kinds of delays into consideration (see Appendix~\ref{subsec:delay}). We denote the set of all simple paths from $s$ to $r$ as $P$. For a path $p\in P$, we denote the number of nodes belonging to $p$ as $|p|$. There are different ways to describe a periodically repeated solution $f$, one of which defines $f$ as the assigned amount of data over $P$ at the time offset $\vec{\mathcal{U}}$,
\be
f ~\triangleq~\left\{ x^p(\vec{u})\ge 0:~~\forall p\in P,\forall \vec{u}\in\vec{\mathcal{U}}\right\},\label{eqn:solution}
\ee
where $\vec{\mathcal{U}}$ is defined as follows: suppose $p\in P$ is an arbitrary path and $p=\langle v_1,v_2,...,v_{|p|}\rangle$, where $\{v_i\in V,i=1,2,...,|p|\}$ are the nodes on $p$ and $\{e_{i-1}=(v_{i-1},v_i)\in E, i=2,3,...,|p|\}$ are the links belonging to $p$, with $v_1=s$ and $v_{|p|}=r$. Any offset $\vec{u}\in \vec{\mathcal{U}}$ corresponding to the path $p$ is described by $\vec{u}=\langle u_0,u_1,u_2,...,u_{|p|}\rangle$, with the following held assuming $u_0=0$ and $d_{e_0}=0$
\be
u_i\in\mathbb{Z}\text{ and }u_i\in[u_{i-1}+d_{e_{i-1}},u_{i-1}+d_{e_{i-1}}+U],~\forall i=1,2,...,|p|.\label{eqn:boundedU}  
\ee  
Each positive $x^p(\vec{u})$ of $f$ requires us to push $x^p(\vec{u})$ amount of data onto link $(v_i,v_{i+1})$ at the offset $u_i$, i.e., push $x^p(\vec{u})$ amount of data of the period that starts at time $k\cdot \mathcal{T}(f)$ onto link $(v_i,v_{i+1})$ at time $k\cdot \mathcal{T}(f)+u_i$, where $\mathcal{T}(f)$ is the task activation period of $f$. We remark that in the definition~\eqref{eqn:boundedU}, we have $u_i-d_{e_{i-1}}-u_{i-1}\le U,\forall i=1,2,...,|p|$. This is equivalent to restricting the data-holding delay of each node to be no more than $U$ slots. Because in this paper we assume each node can hold an arbitrary amount of data at each slot, for our problems $U=+\infty$. However, as proved later in Lem.~\ref{lem:boundedU}, setting $U=T-1$ is large enough for us to solve any feasible instance of our problems, if we are interested in solutions that have a task activation period of $T$. Overall, each positive $x^p(\vec{u})$ of $f$ requires us to transmit $x^p(\vec{u})$ amount of data in a batch from $s$ to $r$, following the path $p$ and the time offset $\vec{u}$.

Given a solution $f$, based on each positive $x^p(\vec{u})$ of $f$, we can easily figure out (i) the beginning offset of pushing those data onto the link $e_i\in p,i=1,...,|p|-1$, denoted as $\mathcal{B}^p(\vec{u},e_i)$,
\be
\mathcal{B}^p(\vec{u},e_i)~=~u_i,\nnb
\ee
and (ii) the end-to-end delay for those data to travel from $s$ to $r$, denoted as $\mathcal{A}^p(\vec{u})$,
\be
\mathcal{A}^p(\vec{u})~=~\mathcal{B}^p(\vec{u},e_{{|p|}-1})+d_{e_{{|p|}-1}}~=~u_{|p|-1}+d_{e_{{|p|}-1}}.\nnb
\ee

One important time-aware networking performance metric of $f$ is the \textbf{maximum delay}, denoted as $\mathcal{M}(f)$. It is the time difference comparing the time when the batch of data of one period is received by the receiver $r$, to the beginning time of this period when those data is generated at the sender $s$ waiting for transmission, i.e.,
\be
\mathcal{M}(f)~\triangleq~\max_{\substack{\forall p\in P,\forall \vec{u}\in\vec{\mathcal{U}}: ~x^p(\vec{u})>0}} \mathcal{A}^p(\vec{u}).\label{eqn:max-delay}
\ee

In order to measure the time that elapsed since the generation of the task period that was most recently delivered to the receiver, we define the \textbf{\textsf{AoI}} of $f$ at time $t$, denoted by $\mathcal{I}(f,t)$, as
\be\label{eqn:age}
\mathcal{I}(f,t)~\triangleq~t-\pi_t(f),
\ee 
where $\pi_t(f)$ is the generation time of the task period that was most recently delivered to $r$ by time $t$, i.e.,
\be
\pi_t(f)~\triangleq~\max_{k\in\mathbb{Z}}\left\{k\cdot \mathcal{T}(f):~k\cdot \mathcal{T}(f)+\mathcal{M}(f)\le t\right\}.\nnb
\ee

\subsection{Problem Definition} 
In this paper we focus on the minimization of (i) the peak value of \textsf{AoI}, and (ii) the average value of \textsf{AoI}, both over all the time slots. We define the \textbf{peak \textsf{AoI}} of $f$, denoted as $\Lambda_p(f)$, as follows
\be
\Lambda_p(f)~\triangleq~\max_{t\in\mathbb{Z}}\mathcal{I}(f,t),\label{eqn:peak-age}
\ee
and define the \textbf{average \textsf{AoI}} of $f$, denoted as $\Lambda_a(f)$, as
\be
\Lambda_a(f)~\triangleq~\sum_{t\in\mathbb{Z}}\mathcal{I}(f,t)~/~\sum_{t\in\mathbb{Z}}1.\label{eqn:avg-age}
\ee         

Our problems of finding a periodically repeated solution $f$ to minimize \textsf{AoI} are subject to a minimum throughput requirement, a maximum throughput requirement, and link bandwidth constraints. The minimum (resp. maximum) \textbf{throughput requirement} requires $f$ to send $D$ amount of data every $\mathcal{T}(f)\in\mathbb{Z}^+$ time slots, achieving a throughput no smaller than an input $R_l\in\mathbb{R}$ (resp. no greater than an input $R_u\in\mathbb{R}$), i.e., 
\be
\sum_{\forall p\in P}\sum_{\forall \vec{u}\in\vec{\mathcal{U}}}x^p(\vec{u})=D, R_l\le D/\mathcal{T}(f)\le R_u,\text{ and }\mathcal{T}(f)\in\mathbb{Z}^+. \label{eqn:throughput}
\ee
It is clear for us to fairly assume $D/R_l\in\mathbb{Z}^+$ and $D/R_u\in\mathbb{Z}^+$ for the input $R_l$ and $R_u$, due to $\mathcal{T}(f)\in\mathbb{Z}^+$.

Given a solution $f$, we denote the aggregate amount of data sent to link $e\in E$ at the offset $i\in\{0,1,...,\mathcal{T}(f)-1\}$, or equivalently the aggregate amount of data sent to $e$ at each time $k\cdot \mathcal{T}(f)+i,\forall k\in\mathbb{Z}$, as $x_e(i)$. Note that $x_e(i)$ may include data assigned to different path-offset pairs of one period, and may even include data from multiple periods with different starting times. We remark that $0\le i\le \mathcal{T}(f)-1,i\in\mathbb{Z}$, because $x_e(i+\mathcal{T}(f))$ is always equal to $x_e(i)$ considering that $f$ is periodically repeated. Specifically, (i) $x_e(i+\mathcal{T}(f))$ is the aggregate data assigned to $e$ at the offset $i+\mathcal{T}(f)$, i.e., at time $k\cdot \mathcal{T}(f)+i+\mathcal{T}(f)$ from the perspective of the period starting at time $k\cdot \mathcal{T}(f)$, and (ii) $x_e(i)$ is the aggregate data assigned to $e$ at the offset $i$, i.e., also at time $k\cdot \mathcal{T}(f)+i+\mathcal{T}(f)$ but from the perspective of the period starting at time $(k+1)\cdot \mathcal{T}(f)$. The link bandwidth constraints require $x_e(i)$ to be no greater than $b_e$, i.e., $x_e(i)\le b_e$, for any link $e\in E$ and any offset $i=0,1,...,\mathcal{T}(f)-1$. This is equivalent to restricting that the aggregate data sent to each link $e\in E$ at each time slot shall be upper bounded by $b_e$.   

It is clear that $x^p(\vec{u})$ will contribute to $x_e(i)$ if and only if $e\in p$ and there exists a $k\in\mathbb{Z}$ such that $k\cdot \mathcal{T}(f)+\mathcal{B}^p(\vec{u},e)=i$. Therefore, our \textbf{link bandwidth constraints} are equivalent to the following
\be
\sum_{\substack{p\in P:\\e\in p}}\sum_{\substack{k\in\mathbb{Z},\vec{u}:k\cdot \mathcal{T}(f)\\+\mathcal{B}^p(\vec{u},e)=i}} x^p(\vec{u})~\le~b_e,~\forall i=0,...,\mathcal{T}(f)-1,\forall e\in E.\label{eqn:capacity}
\ee 

Suppose $\Lambda_{p}^R$ (resp. $\Lambda_a^R$) is the minimal peak \textsf{AoI} (resp. minimal average \textsf{AoI}) that can be achieved by any periodically repeated solution which obtains a throughput of $R$, meeting link bandwidth constraints. Now given a network $G(V,E)$, a sender $s\in V$, a receiver $r\in V\backslash\{s\}$, throughput requirements $R_l$ and $R_u$, in this paper we are interested in the following two \textsf{AoI} minimization problems,
\begin{enumerate}
	\item Obtain an optimal throughput $R_p\in[R_l,R_u],D/R_p\in\mathbb{Z}^+$ that achieves the minimal peak \textsf{AoI}, i.e., 
	\be
	R_p~\triangleq~\arg\min_{R_{l}\le R\le R_{u},D/R\in\mathbb{Z}^+} \Lambda_{p}^R.\nnb
	\ee
	and obtain the feasible periodically repeated solution which has a throughput of $R_p$ and a peak \textsf{AoI} of $\Lambda_p^{R_p}$. We denote this problem of Minimizing Peak \textsf{AoI} as \textbf{\textsf{MPA}}.
	\item Obtain an optimal throughput $R_a\in[R_{l},R_{u}],D/R_a\in\mathbb{Z}^+$ that achieves the minimal average \textsf{AoI}, i.e., 
	\be
	R_a~\triangleq~\arg\min_{R_{l}\le R\le R_{u},D/R\in\mathbb{Z}^+} \Lambda_{a}^R,\nnb
	\ee
	and obtain the feasible periodically repeated solution which has a throughput of $R_a$ and an average \textsf{AoI} of $\Lambda_a^{R_a}$. We denote this problem of Minimizing Average \textsf{AoI} as \textbf{\textsf{MAA}}.
\end{enumerate}

As discussed in Sec.~\ref{sec:related}, existing time-critical multi-path communication problems minimize maximum delay, instead of \textsf{AoI}. Similar to \textsf{MPA} and \textsf{MAA}, we can define (i) $\mathcal{M}^R$ as the minimal maximum delay with a throughput of $R$, and (ii) problem of Minimizing Maximum Delay (\textbf{\textsf{MMD}}) as the problem of obtaining an optimal $R_m\in[R_{l},R_{u}],D/R_m\in\mathbb{Z}^+$ that achieves minimal maximum delay, and obtaining associated optimal periodically repeated solution.

Finally, we give one lemma which argues for any feasible solution $g$ whose data-holding delay may exceed $\mathcal{T}(g)-1$ slots for certain node, there must exist a feasible solution $f$ whose data-holding delay is no more than $\mathcal{T}(f)-1$ slots for all nodes, and the following holds comparing $g$ with $f$: (i) they achieve the same throughput, and (ii) the peak \textsf{AoI} (resp. average \textsf{AoI}) of $f$ is no worse than that of $g$. A direct corollary is for any feasible \textsf{MPA} (resp. \textsf{MAA}) instance, setting $U$ (see formula~\eqref{eqn:solution}) to be $T-1$ is large enough for us to solve it, if we are interested in solutions which have a task activation period of $T$ and thus a throughput of $D/T$. 
\begin{mdframed}[skipabove=10pt, skipbelow=10pt, roundcorner=10pt, linewidth=0pt, backgroundcolor=gray!10]
\begin{lemma}\label{lem:boundedU}
	Given any instance of \textsf{MPA} (or \textsf{MAA}), suppose $g$ is an arbitrary feasible periodically repeated solution. Then there must exist another feasible periodically repeated solution $f$, where $\mathcal{T}(f)=\mathcal{T}(g)$, $\Lambda_p(f)\le \Lambda_p(g)$, $\Lambda_a(f)\le \Lambda_a(g)$, and for each positive $x^p(\vec{u})$ (suppose $p=\langle v_1,...,v_{|p|}\rangle$ and $\vec{u}=\langle u_0,...,u_{|p|}\rangle$) of $f$, we have $u_i-d_{e_{i-1}}-u_{i-1}\le \mathcal{T}(f)-1$, for all $i=1,2,...,|p|$.
\end{lemma}
\end{mdframed}
\begin{proof}
	Refer to Appendix~\ref{subsec:boundedU}.
\end{proof}
\section{Compare \textsf{AoI} with Maximum Delay}\label{sec:delay-aoi}
As time-critical networking performance metrics, maximum delay is well-known, while \textsf{AoI} is newly proposed. In this section, we compare the problem of minimizing \textsf{AoI} (\textsf{MPA} and \textsf{MAA}) with that of minimizing maximum delay (\textsf{MMD}) theoretically.


Consider the following example. In a network with nodes $s$ and $r$, and one link $(s,r)$. Suppose the delay (resp. bandwidth) of the link is $d$ (resp. $b\ge D$). Suppose $R_u=D$ and $R_l=D/T_u$ given a $T_u\in\mathbb{Z}^+$. Consider one solution that streams $D$ data to $(s,r)$ at the offset $0$. It is clear that this solution can have a task activation period of $T\le T_u$, meeting throughput requirements and link bandwidth constraints. And the batch of data of the period starting at time $kT$ will be received by $r$ at time $kT+d$. Now consider two different task activation periods $T_1$ and $T_2$ with $T_1<T_2\le T_u$. From the perspective of minimizing maximum delay, the solution with $T=T_1$ is equivalent to that with $T=T_2$, because they are both feasible, and obtain the same maximum delay of $d$. From the perspective of minimizing peak/average \textsf{AoI}, in contrast, the solution with $T=T_1$ is better than that with $T=T_2$, since according to Lem.~\ref{lem:age-delay} introduced later, the peak \textsf{AoI} (resp. average \textsf{AoI}) of former is $d+T_1-1$ (resp. $d+(T_1-1)/2$), which is smaller than that of latter, i.e., than $d+T_2-1$ (resp. $d+(T_2-1)/2$). In fact, $T_1$ is better than $T_2$ in this example, because they lead to the same delay of periodically transmitting the batch of data, but the throughput achieved by $T_1$ ($D/T_1$) is greater than that achieved by $T_2$ ($D/T_2$).

For periodic transmission services, \textsf{AoI}, instead of maximum delay, should be optimized to provide time-critical solutions according to the example. This is mainly because \textsf{AoI} is a time-critical metric simultaneously considering throughput and maximum delay. In the following, we further prove that the maximum-delay-optimal solution can achieve a suboptimal peak/average \textsf{AoI}, but it must be with bounded optimality loss compared to optimal.

Given a solution $f$, first we give a lemma to mathematically relates the peak/average \textsf{AoI} of $f$ to the maximum delay of $f$.
\begin{mdframed}[skipabove=10pt, skipbelow=10pt, roundcorner=10pt, linewidth=0pt, backgroundcolor=gray!10]
\begin{lemma}\label{lem:age-delay}
	For an arbitrary periodically repeated solution $f$, we have the following
	\be
	\Lambda_p(f)~=~\mathcal{M}(f)+\mathcal{T}(f)-1,~~\Lambda_a(f)~=~\mathcal{M}(f)+(\mathcal{T}(f)-1)/2.\nnb
	\ee
\end{lemma}
\end{mdframed}
\begin{proof}
	Refer to Appendix~\ref{subsec:age-delay}.
\end{proof}

A direct corollary is that the peak \textsf{AoI} (resp. average \textsf{AoI}) of a feasible solution which achieves a throughput of $R$ and has a maximum delay of $\mathcal{M}^R$ is $\Lambda_{p}^R$ (resp. $\Lambda_{a}^R$). Thus to solve \textsf{MPA} and \textsf{MAA} given $R_{l}=R_{u}$, we can solve the corresponding \textsf{MMD} instead. However, as introduced in Sec.~\ref{sec:related}, only special cases of \textsf{MMD} with $R_{l}=R_{u}$, i.e., the quickest flow problem ($R_{l}=R_{u}\to 0$) and the min-max-delay flow problem ($R_{l}=R_{u}=D$), are studied in the literature, and it is not clear how to solve \textsf{MMD} even given $0< R_{l}=R_{u}\le D$. Moreover, for general settings with $R_l< R_u$, we observe that both \textsf{MPA} and \textsf{MAA} can differ from \textsf{MMD} as follows.
\begin{mdframed}[skipabove=10pt, skipbelow=10pt, roundcorner=10pt, linewidth=0pt, backgroundcolor=gray!10]
\begin{lemma}\label{lem:delay-aoi}
	Given any instance of \textsf{MPA} (or \textsf{MAA}, \textsf{MMD}), suppose $\vec{R}_p$ (resp. $\vec{R}_a$, $\vec{R}_m$) is the optimal set of throughputs that minimize peak \textsf{AoI} (resp. average \textsf{AoI}, maximum delay) of this instance. The following must hold for this instance
	\be
	\min_{R_p\in \vec{R}_p}R_p~\ge~ \max_{R_m\in \vec{R}_m}R_m,~~\min_{R_a\in \vec{R}_a}R_a~\ge~ \max_{R_m\in \vec{R}_m}R_m.\nnb
	\ee
	And there must exist an instance where the following holds
	\be
	\min_{R_p\in \vec{R}_p}R_p~>~ \max_{R_m\in \vec{R}_m}R_m,~~\min_{R_a\in \vec{R}_a}R_a~>~ \max_{R_m\in \vec{R}_m}R_m.\nnb
	\ee
\end{lemma}
\end{mdframed}
\begin{proof}
	Refer to Appendix~\ref{subsec:delay-aoi}.
\end{proof}

In Lem.~\ref{lem:delay-aoi}, $\vec{R}_p$ is defined as a set of throughputs, because in certain instances there may exist multiple throughputs obtaining the same and optimal peak \textsf{AoI}. Similarly, we define $\vec{R}_a$ and $\vec{R}_m$ both as sets of throughputs.

Lem.~\ref{lem:delay-aoi} suggests that (i) minimizing maximum delay can differ from minimizing \textsf{AoI}, because the maximum-delay-optimal solution can achieve suboptimal peak/average \textsf{AoI}. (ii) The throughput of the maximum-delay-optimal solution must be no greater than that of the peak-/average- \textsf{AoI}-optimal solution. In the following, we further characterize near-tight optimality losses for the suboptimal \textsf{AoI} achieved by the maximum-delay-optimal solution.
\begin{mdframed}[skipabove=10pt, skipbelow=10pt, roundcorner=10pt, linewidth=0pt, backgroundcolor=gray!10]
\begin{lemma}\label{lem:delay-aoi-gap}
	Given any instance of \textsf{MPA} (or \textsf{MAA}, \textsf{MMD}), suppose $\vec{R}_p$ (resp. $\vec{R}_a$, $\vec{R}_m$) is the optimal set of throughputs that minimize peak \textsf{AoI} (resp. average \textsf{AoI}, maximum delay) of this instance. The following must hold for this instance
	\bee
	&& \Lambda_{p}^{R_m}-\Lambda_{p}^{R_p}\le \frac{D}{R_l}-\frac{D}{R_u},\forall R_m\in\vec{R}_m,\forall R_p\in\vec{R}_p.\label{eqn:delay-peak-gap}\\
	&& \Lambda_{a}^{R_m}-\Lambda_{a}^{R_a}\le \frac{D}{2R_l}-\frac{D}{2R_u},\forall R_m\in\vec{R}_m,\forall R_a\in\vec{R}_a\label{eqn:delay-avg-gap}	
	\eee	
	Gap~\eqref{eqn:delay-peak-gap} is near-tight, in the sense that for arbitrary $D$, $R_l$, and $R_u$ that meet $D>0$, $D/R_{l}\in\mathbb{Z}^+$, and $D/R_{u}\in\mathbb{Z}^+$, there is an instance where the following holds
	\be
	\Lambda_{p}^{R_m}-\Lambda_{p}^{R_p}~\ge~\frac{D}{R_l}-\frac{D}{R_u}-1,~~\forall R_m\in\vec{R}_m,\forall R_p\in\vec{R}_p.\nnb
	\ee
	Gap~\eqref{eqn:delay-avg-gap} is near-tight, in a similar sense with the following held
	\be
	\Lambda_{a}^{R_m}-\Lambda_{a}^{R_a}~\ge~\frac{D}{2R_l}-\frac{D}{2R_u}-1,~~\forall R_m\in\vec{R}_m,\forall R_a\in\vec{R}_a.\nnb
	\ee
\end{lemma}
\end{mdframed}
\begin{proof}
	Refer to Appendix~\ref{subsec:delay-aoi-gap}.
\end{proof}

Overall, we observe that \textsf{MPA} and \textsf{MAA} are non-trivial as compared to \textsf{MMD}: (i) \textsf{AoI}-optimal solution, instead of maximum-delay-optimal one, is the time-critical solution for periodic transmission services; (ii) \textsf{AoI}-optimal solution can differ from the maximum-delay-optimal one in the general scenario with throughput optimization involved ($R_{l}< R_{u}$); (iii) even for the special scenario where the throughput of feasible solutions is fixed ($R_{l}=R_{u}$), where it can be proved that the \textsf{AoI}-optimal solution is also maximum-delay-optimal, and vice versa, existing maximum delay minimization studies have strong assumptions on the fixed throughput (either $R_{l}=R_{u}\to 0$ or $R_{l}=R_{u}=D$), and it is not clear how to minimize maximum delay with the throughput fixed arbitrarily ($0<R_{l}=R_{u}\le D$). In the following sections, we design an optimal algorithm and an approximation framework for \textsf{MPA} and \textsf{MAA}.
\section{Problem Structures of \textsf{MPA} and \textsf{MAA}}\label{sec:pseudo}
In this section we give a complete understanding on the fundamental structures of our \textsf{MPA} and \textsf{MAA}. In particular, we first show that \textsf{MPA} and \textsf{MAA} are two different problems theoretically, and then prove that they are both NP-hard in the weak sense, with a pseudo-polynomial-time optimal algorithm developed.
\subsection{\textsf{MPA} is Different from \textsf{MAA}}
Comparing \textsf{MPA} of minimizing peak \textsf{AoI} with \textsf{MAA} of minimizing average \textsf{AoI}, we observe that they are two different problems, as proved in the following lemma.

\begin{mdframed}[skipabove=10pt, skipbelow=10pt, roundcorner=10pt, linewidth=0pt, backgroundcolor=gray!10]
	\begin{lemma}\label{lem:peak-average}
		Given any instance of \textsf{MPA} (or \textsf{MAA}), suppose $\vec{R}_p$ (resp. $\vec{R}_a$) is the optimal set of throughputs that minimize peak \textsf{AoI} (resp. average \textsf{AoI}) of this instance. For this instance,
		\begin{enumerate}
	        \item the following must hold
			\be
			\min_{R_p\in \vec{R}_p}R_p\ge \max_{R_a\in \vec{R}_a}R_a,~~\min_{R_p\in \vec{R}_p}\mathcal{M}^{R_p}\ge\max_{R_a\in \vec{R}_a}\mathcal{M}^{R_a},\nnb
			\ee
			\item and we have the following 
			\bee
			&& \Lambda_{a}^{R_p}-\Lambda_{a}^{R_a}\le\frac{D}{2R_l}-\frac{D}{2R_u},\forall R_p\in\vec{R}_p,\forall R_a\in\vec{R}_a.\label{eqn:gap-1}\\
			&& \Lambda_{p}^{R_a}-\Lambda_{p}^{R_p}\le\left\lfloor\frac{D}{2R_l}-\frac{D}{2R_u}\right\rfloor,\forall R_p\in\vec{R}_p,\forall R_a\in\vec{R}_a.\label{eqn:gap-2}	
			\eee 
		\end{enumerate}
		Moreover, there must exist an instance where the following holds
		\be
		\min_{R_p\in \vec{R}_p}R_p> \max_{R_a\in \vec{R}_a}R_a,~~\min_{R_p\in \vec{R}_p}\mathcal{M}^{R_p}>\max_{R_a\in \vec{R}_a}\mathcal{M}^{R_a},\nnb
		\ee
	\end{lemma}
\end{mdframed}
\begin{proof}
	Refer to Appendix~\ref{subsec:peak-average}.
\end{proof}

From the lemma, we learn that (i) \textsf{MPA} can differ from \textsf{MAA}, and (ii) although both \textsf{MPA} and \textsf{MAA} minimize \textsf{AoI} which jointly considers throughput and maximum delay, we observe that \textsf{MPA} of minimizing peak \textsf{AoI} may carry more flavor on throughput and less on maximum delay, compared to \textsf{MAA} of minimizing average \textsf{AoI}. In the lemma, (iii) we further characterize bounded optimality loss for the suboptimal average \textsf{AoI} (resp. suboptimal peak \textsf{AoI}) achieved by the optimal solution to \textsf{MPA} (resp. to \textsf{MAA}). 
\subsection{\textsf{MPA} and \textsf{MAA} are both NP-Hard}
Although \textsf{MPA} differs from \textsf{MAA}, we observe that they are both NP-hard, because (i) based on Lem.~\ref{lem:age-delay}, \textsf{MMD} given $R_{l}=R_{u}=D$ is a special case of \textsf{MPA} and \textsf{MAA}. (ii) As discussed in Sec.~\ref{sec:related}, the min-max-delay flow problem under our setting is exactly the problem \textsf{MMD} given $R_{l}=R_{u}=D$, and it has been proven to be NP-hard by the study~\cite{misra2009polynomial}. Overall, we have the following.

\begin{mdframed}[skipabove=10pt, skipbelow=10pt, roundcorner=10pt, linewidth=0pt, backgroundcolor=gray!10]
\begin{lemma}\label{lem:np-hard}
	\textsf{MPA} and \textsf{MAA} are both NP-hard.
\end{lemma}
\end{mdframed}
\begin{proof}
	Both \textsf{MPA} and \textsf{MAA} cover the NP-hard min-max-delay flow problem~\cite{misra2009polynomial} as a special case.
\end{proof}

In the following we propose a pseudo-polynomial-time algorithm which solves \textsf{MPA} (resp. \textsf{MAA}) optimally. It enumerates all possible throughputs $R\in[R_{l},R_{u}],D/R\in\mathbb{Z}^+$ to figure out the peak-\textsf{AoI}-optimal $R_p$ (resp. average-\textsf{AoI}-optimal $R_a$), together with the optimal periodically repeated solution.

\subsection{Design an Algorithm~\ref{alg:pseudo} to Obtain $\textsf{M}^R$ in a Pseudo-Polynomial Time}
Given a throughput $R$ with $D/R\in\mathbb{Z}^+$, first we design a pseudo-polynomial-time algorithm which leverages a binary-search based scheme, together with an expanded network, to figure out the minimal maximum delay $\mathcal{M}^R$ and the corresponding solution. According to Lem.~\ref{lem:age-delay}, the minimal peak \textsf{AoI} $\Lambda_{p}^R$ and the minimal average \textsf{AoI} $\Lambda_{a}^R$ can be achieved by the same solution.

\begin{figure}[]
	\centering
	\vspace{-2ex}
	\includegraphics[width=0.9\columnwidth]{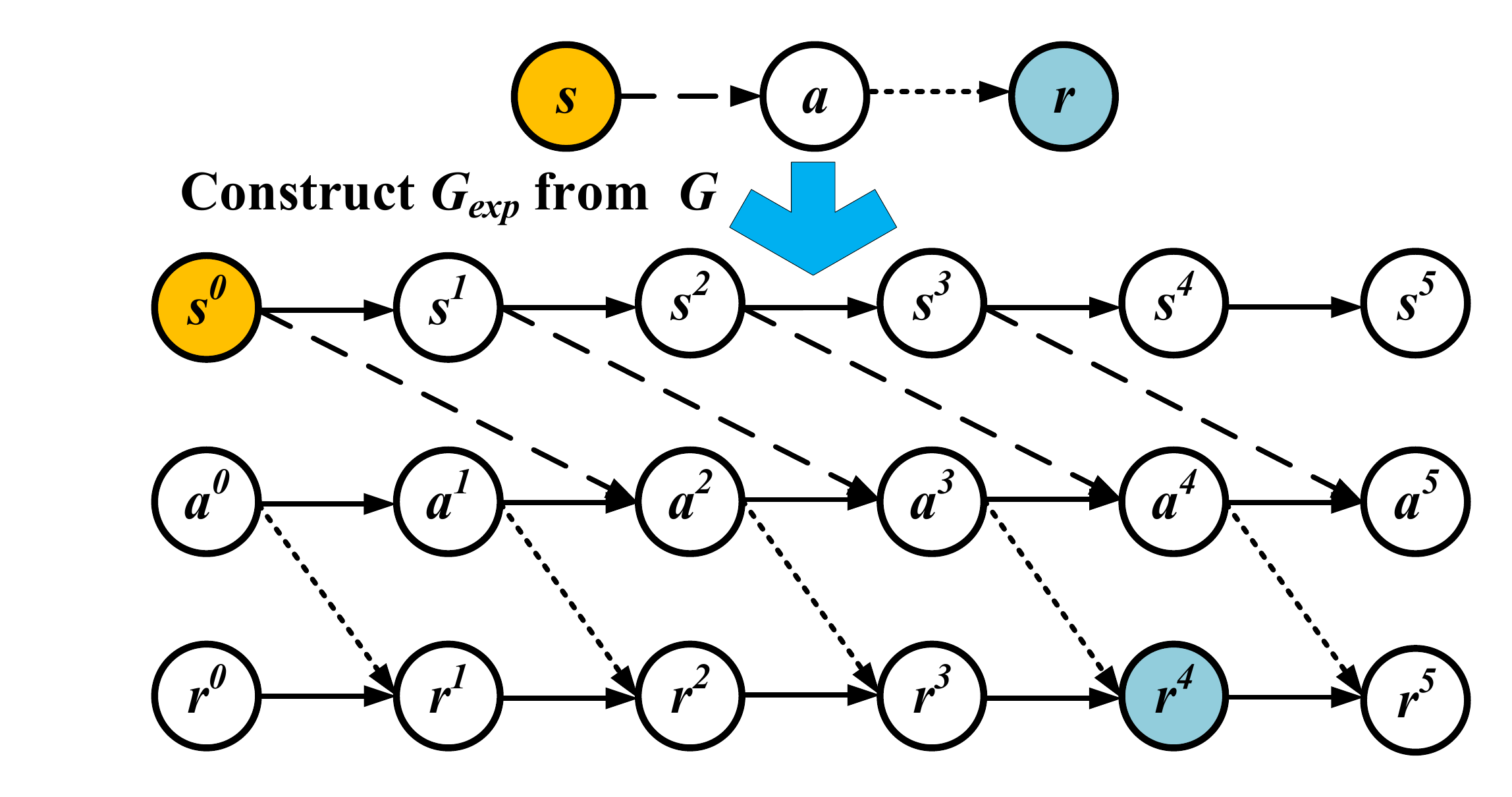}
	\vspace{-2ex}
	\caption{An example of constructing $G_{\textsf{exp}}$. Suppose $d_{(s,a)}=2$ and $d_{(a,r)}=1$ in $G$. And suppose $M_U=5$ and $M=4$.}
	\label{fig:expanded-graph}
\end{figure}

\textbf{Construct an expanded network}. We construct an expanded network $G_{\textsf{exp}}(V_{\textsf{exp}},E_{\textsf{exp}})$, from the input $G(V,E)$ following Algorithm~\ref{alg:expanded-graph}. Given an integer $M_U$ that is an upper bound of $\mathcal{M}^R$, first we expand each node $v\in V$ to nodes $v^i,i=0,...,M_U$ (the loop in line~\ref{line:expand-node}). By this expansion, node $v^i$ represents the node $v$ at time $kT+i$ from the perspective of the period starting at time $kT,\forall k\in\mathbb{Z}$, where $T=D/R$. Second we expand each link $e=(v,w)\in E$ to links $(v^i,w^{i+d_e}),i=0,...,M_U-d_e$ (the loop in line~\ref{line:expand-link}). By this expansion, the link $(v^i,w^{i+d_e})$ represents that certain amount of data can be streamed to the link $(v,w)$ at time $kT+i$ from the perspective of the period starting at time $kT,\forall k\in\mathbb{Z}$, where $T=D/R$. Third, we add links $(v^i,v^{i+1}),i=0,...,M_U-1$ (the loop in line~\ref{line:r-link}) for each $v\in V$, because we allow each node to hold data at each time slot.

\begin{algorithm}
	\caption{Construct $G_{\textsf{exp}}$ from $G$}\label{alg:expanded-graph}
	\begin{algorithmic}[1]
		\State \textbf{input}: $G=(V,E)$, $M_U$
		\State \textbf{output}: $G_{\textsf{exp}}=(V_{\textsf{exp}},E_{\textsf{exp}})$
		\Procedure{}{}
		\State $V_{\textsf{exp}}=E_{\textsf{exp}}=\textsf{NULL}$
		\For{$v\in V$ and $i=0,1,...,M_U$}\label{line:expand-node}
		\State Push node $v^i$ into $V_{\textsf{exp}}$ 
		\EndFor
		\For{$e=(v,w)\in E$ and $i=0,1,...,M_U-d_e$}\label{line:expand-link}
		\State Push link $(v^i,w^{i+d_e})$ into $E_{\textsf{exp}}$ 
		\EndFor
		\For{$v\in V$ and $i=0,1,...,M_U-1$}\label{line:r-link}
		\State Push link $(v^i,v^{i+1})$ into $E_{\textsf{exp}}$ 
		\EndFor
		\State \textbf{return} $G_{\textsf{exp}}=(V_{\textsf{exp}},E_{\textsf{exp}})$
		\EndProcedure
	\end{algorithmic}
\end{algorithm}

\textbf{Obtain $\mathcal{M}^R$ using binary search}. Given an arbitrary integer $M$ with $M\le M_U$, we observe that the problem of whether there exists a feasible periodically repeated solution $f$ in $G$, with $\mathcal{T}(f)=D/R$ and $\mathcal{M}(f)\le M$, can be solved by solving a network flow problem that is casted by the following linear program in $G_{\textsf{exp}}$.
\bse
\label{eqn:Expanded}
\bee
\max && \sum_{e'\in \textsf{Out}(s^0)} x_{e'}\label{eqn:objective}
\\ \mbox{s.t. } && \sum_{e'\in \textsf{Out}(s^0)} x_{e'} =\sum_{e'\in \textsf{In}(r^M)} x_{e'},\label{eqn:conservation1}\\
&& \sum_{e'\in \textsf{Out}(v)} x_{e'} =\sum_{e'\in \textsf{In}(v)} x_{e'},\forall v\in V_{\textsf{exp}}\backslash\{s^0,r^M\},\label{eqn:conservation2}\\
&& \sum_{e'\in e(i)} x_{e'}\le b_e,~\forall e\in E, \forall i=0,1,...,D/R-1,\label{eqn:capacity-expanded}\\
\mbox{vars. } && x_{e'}\ge 0,~\forall e'\in E_{\textsf{exp}}.
\eee
\ese
Here $\textsf{In}(v)$ (resp. $\textsf{Out}(v)$) is the set of incoming (resp. outgoing) links of $v\in V_{\textsf{exp}}$ in $G_{\textsf{exp}}$. Suppose $e=(v,w)\in E$, then $e(i)$ is the set of expanded links $\{(v^{kT+i},w^{kT+i+d_e}),\forall k\in\mathbb{Z}\}$ that belong to $E_{\textsf{exp}}$, where $T=D/R$. Note that data assigned to $e(i)$ must aggregately respect bandwidth constraint of $b_e$, considering that the aggregate data assigned to $e(i)$ is exactly equal to $x_e(i)$ that is introduced in the definition of our link bandwidth constraints~\eqref{eqn:capacity}. This is because the difference of starting times of links belonging to $e(i)$ are multiples of the task activation period $D/R$. The objective~\eqref{eqn:objective} maximizes the amount of data sent from the sender of each period. Constraint~\eqref{eqn:conservation1} restricts those data arrive at the receiver no later than $M$ time slots as compared to the starting time of the period. Constraints~\eqref{eqn:conservation2} are flow conservation constraints, and constraints~\eqref{eqn:capacity-expanded} are link bandwidth constraints. We remark again that the constraints~\eqref{eqn:capacity-expanded} restricts that the aggregate data pushed onto $e$ at each time $kT+i,\forall k\in\mathbb{Z}$ shall be upper bounded by the bandwidth $b_e$, for all $e\in E$ and all $i=0,1,...,T-1$, where $T=D/R$.


\begin{mdframed}[skipabove=10pt, skipbelow=10pt, roundcorner=10pt, linewidth=0pt, backgroundcolor=gray!10]
\begin{lemma}\label{lem:expanded}
	Given any instance of \textsf{MPA} (or \textsf{MAA}, \textsf{MMD}), suppose $R$ is an arbitrary throughput satisfying $R\in[R_{l},R_{u}]$ and $D/R\in\mathbb{Z}^+$. Let us assume $M$ to be an arbitrary integer. Then the problem of whether there exists a feasible periodically repeated solution $f$ with $\mathcal{T}(f)=D/R$ and $\mathcal{M}(f)\le M$ is feasible if and only if the value of the optimal solution to the linear program~\eqref{eqn:Expanded} is no smaller than $D$.
\end{lemma}
\end{mdframed}
\begin{proof}
	Refer to Appendix~\ref{subsec:expanded}.
\end{proof}

To obtain $\mathcal{M}^R$, Lem.~\ref{lem:expanded} suggests that we can use binary search to obtain the minimal integer $M^*\in[0,M_U]$, under which the linear program~\eqref{eqn:Expanded} outputs a feasible flow with a value no smaller than $D$, and it is clear that the achieved $M^*$ shall be the $\mathcal{M}^R$ (see Algorithm~\ref{alg:pseudo}). Note that to construct the expanded network, we need a $M_U\ge \mathcal{M}^R$. We remark that $M_U$ must exist, e.g., we can set $M_U=|V|\cdot (d_{\max}+D/R_l)$ with $d_{\max}=\max_{e\in E}d_e$, since for any path $p\in P$ and any offset $\vec{u}\in\vec{\mathcal{U}}$ that corresponds to $p$, the following holds for any periodically repeated solution: (i) $|V|\cdot d_{\max}$ is an upper bound of the aggregate delay experienced by passing all the links that belong to $p$, since $p$ is simple, and (ii) $|V|D/R_l$ is an upper bound of the aggregate data-holding delay at all the nodes that belong to $p$, due to our Lem.~\ref{lem:boundedU}.
\begin{algorithm}
	\caption{Obtain $\mathcal{M}^R$ and the corresponding solution}\label{alg:pseudo}
	\begin{algorithmic}[1]
		\State \textbf{input}: $G=(V,E)$, $R$, $D$, $M_U$, $s$, $r$
		\State \textbf{output}: $f$, $\mathcal{M}$
		\Procedure{}{}
		\State $f=f_t=\textsf{NULL}$, $\mathcal{M}=+\infty$, $LB=0$, $UB=M_U$
		\State Obtain $G_{\textsf{exp}}$ by Algorithm~\ref{alg:expanded-graph} with $(G,M_U)$
		\While{$LB\le UB$}
		\State $M=\lceil (LB+UB)/2\rceil$
		\State \parbox[t]{\dimexpr\linewidth-\algorithmicindent}{$f_t$ is the solution by solving the linear program~\eqref{eqn:Expanded} with input $(G_{\textsf{exp}},R,D,M,s,r)$}
		\If{the objective of $f_t$ is no smaller than $D$}
		\State $f=f_t$, $\mathcal{M}=M$, $UB=M-1$
		\Else
		\State $LB=M+1$
		\EndIf
		\EndWhile
		\State \textbf{return} $f$, $\mathcal{M}$
		\EndProcedure
	\end{algorithmic}
\end{algorithm}
\begin{mdframed}[skipabove=10pt, skipbelow=10pt, roundcorner=10pt, linewidth=0pt, backgroundcolor=gray!10]
\begin{lemma}\label{lem:time-complexity}
	Suppose $\mathcal{L}$ is the input size of the instance of linear program~\eqref{eqn:Expanded}, then the time complexity of Algorithm~\ref{alg:pseudo} is $O(|E|^3M_U^3\mathcal{L}\log M_U)$.
\end{lemma}
\end{mdframed}
\begin{proof}
	Refer to Appendix~\ref{subsec:time-complexity}.
\end{proof}

Lem.~\ref{lem:time-complexity} shows that our Algorithm~\ref{alg:pseudo} has a pseudo-polynomial time complexity, because of the pseudo-polynomial size of the expanded network: (i) considering $M_U\le |V|\cdot (d_{\max}+D/R_l)$, the time complexity is polynomial with the numeric value of $d_{\max}$ and $D/R_l$, but (ii) it is exponential with the bit length of $d_{\max}$ and $D/R_l$.
\subsection{Use Algorithm~\ref{alg:pseudo} to Solve \textsf{MPA} and \textsf{MAA} Optimally in a Pseudo-Polynomial Time}\label{subsec:pseudo-algorithm}
We remark that it is challenging to obtain the optimal throughput $R_p\in[R_{l},R_{u}]$ (resp. $R_a\in[R_{l},R_{u}]$) that minimizes peak \textsf{AoI} (resp. average \textsf{AoI}), due to the following observation.
\begin{mdframed}[skipabove=10pt, skipbelow=10pt, roundcorner=10pt, linewidth=0pt, backgroundcolor=gray!10]
\begin{lemma}\label{thm:age-T}
		Both $\Lambda_{p}^R$ and $\Lambda_{a}^R$ are non-monotonic, non-convex, and non-concave with $R$ theoretically.
\end{lemma}
\end{mdframed}
\begin{proof}
	Refer to Appendix~\ref{sub:age-T}.
\end{proof}

Thus to solve \textsf{MPA} (resp. \textsf{MAA}) optimally, we need to enumerate $\Lambda_{p}^R$ (resp. $\Lambda_{a}^R$) for all $R\in[R_{l},R_{u}],D/R\in\mathbb{Z}^+$, and obtain the optimal one that achieves minimal peak \textsf{AoI} (resp. minimal average \textsf{AoI}). It is clear that we can use Algorithm~\ref{alg:pseudo} to achieve $\Lambda_{p}^R$ and $\Lambda_{a}^R$. Therefore, we suggest to solve \textsf{MPA} (resp. \textsf{MAA}) optimally using Algorithm~\ref{alg:pseudo} by enumerating all possible throughputs.

We remark that our proposed enumerating approach has a pseudo-polynomial time complexity. As shown in Lem.~\ref{lem:time-complexity}, Algorithm~\ref{alg:pseudo} has a pseudo-polynomial time complexity to obtain $\Lambda_{p}^R$ and $\Lambda_{a}^R$. Now considering that the number of the enumerated throughputs is $D/R_{l}-D/R_{u}+1$ which is pseudo-polynomial with $D/R_l$, using Algorithm~\ref{alg:pseudo} to solve \textsf{MPA} and \textsf{MAA} optimally by enumeration has a pseudo-polynomial time complexity, too.

Overall, we have the following theorem for \textsf{MPA} and \textsf{MAA}.
\begin{mdframed}[skipabove=10pt, skipbelow=10pt, roundcorner=10pt, linewidth=0pt, backgroundcolor=gray!10]
\begin{theorem}\label{thm:weak-NP-hard}
	\textsf{MPA} and \textsf{MAA} are NP-hard in the weak sense.
\end{theorem}
\end{mdframed}
\begin{proof}
	It is a direct result from Lem.~\ref{lem:np-hard} and our proposed optimal algorithm which has a pseudo-polynomial time complexity.
\end{proof}
\section{An Approximation Framework}\label{sec:approximation}
As discussed in Sec.~\ref{sec:delay-aoi}, the peak/average \textsf{AoI} of the solution minimizing maximum delay is within a bounded gap as compared to optimal. Thus it is natural to use approximate solutions to the problem of minimizing maximum delay as approximate solutions to our problems of minimizing \textsf{AoI}. However, this idea is non-trivial, considering that as discussed in Sec.~\ref{sec:related}, existing maximum delay minimization problems (i.e., the quickest flow problem and the min-max-delay flow problem) are just special cases of the maximum-delay-minimization counterpart of our \textsf{AoI} minimization problems. This is because they assume a fixed task activation period, which is quite different from our problems that assume the task activation period to be decision variables. In this section, we overcome the challenge, and propose a framework that can adapt any polynomial-time approximation algorithm of the min-max-delay flow problem to solve our \textsf{MPA} and \textsf{MAA} approximately in a polynomial time.


For any feasible periodically repeated solution $f$ to \textsf{MPA} and \textsf{MAA} achieving a throughput of $R$, it should send $D$ amount of data from $s$ to $G$ every $D/R$ slots, meeting link bandwidth constraints. According to the definition of the min-max-delay flow problem (refer to~\cite{misra2009polynomial}), for any feasible solution $\textbf{f}$ to the min-max-delay flow problem achieving a throughput of $R$, it should send $R$ amount of data from $s$ to $G$ at each slot, meeting link bandwidth constraints. Because it is clear that this $\textbf{f}$ can send $D$ amount of data from $s$ to $G$ every $D/R$ slots, meeting link bandwidth constraints, we observe that $\textbf{f}$ is a special case of $f$. 

Let us denote a feasible instance of \textsf{MPA} (resp. \textsf{MAA}) characterized by $(G,s,r,R_{l},R_{u},D)$ as \textbf{$\textsf{MPA}(R_l,R_u, D)$} (resp. \textbf{$\textsf{MAA}(R_l,R_u, D)$}). And denote the corresponding min-max-delay flow problem instance, which is defined by the same $G$, $s$, $r$, but with a throughput requirement of $R$, as \textbf{$\textsf{MMD1}(R)$} (note as discussed in Sec.~\ref{sec:related}, min-max-delay flow problem assumes a fixed task activation period of $1$, and thus a fixed throughput requirement, but \textsf{MPA} and \textsf{MAA} assume both a minimum and maximum throughput requirement). We have the following lemma.

\begin{mdframed}[skipabove=10pt, skipbelow=10pt, roundcorner=10pt, linewidth=0pt, backgroundcolor=gray!10]
\begin{lemma}\label{lem:MLP}
	Given any $\textsf{MPA}(R_l,R_u, D)$ (resp. $\textsf{MAA}(R_l,R_u, D)$), suppose $R\in[R_{l},R_{u}],D/R\in\mathbb{Z}^+$ is an arbitrary feasible throughput for it. Then $\textsf{MMD1}(R)$ must be feasible. Moreover, suppose $\textbf{f}(R)$ is an arbitrary feasible solution to $\textsf{MMD1}(R)$, it holds that $\textbf{f}(R)$ must be a feasible periodically repeated solution to $\textsf{MPA}(R_l,R_u, D)$ (resp. $\textsf{MAA}(R_l,R_u, D)$) with the following
	\be
	\mathcal{M}(\textbf{f}(R))~=~\hat{\mathcal{M}}(\textbf{f}(R))+D/R-1,\nnb
	\ee
	where $\hat{\mathcal{M}}(\textbf{f})$ is the maximum delay of $\textbf{f}$ with $\textsf{MMD1}(R)$.
\end{lemma}
\end{mdframed}
\begin{proof}
	Referred to Appendix~\ref{subsec:MLP}.
\end{proof}

Lem.~\ref{lem:MLP} suggests that any feasible solution to the min-max-delay flow problem achieving a throughput of $R$ is a feasible periodically repeated solution to the corresponding \textsf{MPA} and \textsf{MAA} also achieving a throughput of $R$. But we remark that even for the optimal solution to the min-max-delay flow problem, its peak \textsf{AoI} (resp. average \textsf{AoI}) can be strictly greater than the minimal peak \textsf{AoI} (resp. minimal average \textsf{AoI}) with a throughput of $R$, i.e., than $\Lambda_{p}^R$ (resp. $\Lambda_{a}^R$). This is because when we look at a solution to the min-max-delay flow problem from the perspective of \textsf{MPA} and \textsf{MAA}, it always sends $R$ amount of data from $s$ to $G$ at each slot, which is a special case of feasible solutions to \textsf{MPA} and \textsf{MAA}. In fact, \textsf{MPA} and \textsf{MAA} allow various amount of data to be sent to $G$ at each slot, as long as a total of $D$ amount of data can be sent every $D/R$ slots.

Lem.~\ref{lem:MLP} suggests that we can use the solution to the min-max-delay flow problem as a solution to our \textsf{MPA} (resp. \textsf{MAA}). As it is easy to prove that if $\textsf{MMD1}(R_1)$ is feasible, $\textsf{MMD1}(R_2)$ must be feasible given any $0<R_2\le R_1$ (see the proof to the following theorem), a direct result of Lem.~\ref{lem:MLP} is that $R_l$ must be a feasible throughput for $\textsf{MPA}(R_l,R_u,D)$ (resp. $\textsf{MAA}(R_l,R_u,D)$). Therefore, it is clear that solving $\textsf{MMD1}(R_l)$ must output a feasible solution to $\textsf{MPA}(R_l,R_u,D)$ (resp. $\textsf{MAA}(R_l,R_u,D)$). In the following theorem, we further prove that any approximate solution to $\textsf{MMD1}(R_l)$ must be an approximate solution to $\textsf{MPA}(R_l,R_u,D)$ (resp. $\textsf{MAA}(R_l,R_u,D)$), with bounded optimality loss. For easier reference, we denote an arbitrary $\alpha$-approximation algorithm of the min-max-delay flow problem as \textbf{$\textsf{ALG-MMD1}(\alpha)$}.

\begin{mdframed}[skipabove=10pt, skipbelow=10pt, roundcorner=10pt, linewidth=0pt, backgroundcolor=gray!10]
\begin{theorem}\label{thm:MLP-AOI}
	Given any $\textsf{MPA}(R_l,R_u,D)$ and $\textsf{MAA}(R_l,R_u,D)$ where $D/R_l\in\mathbb{Z}^+$, $D/R_u\in\mathbb{Z}^+$, suppose we use $\textsf{ALG-MMD1}(\alpha)$ to solve the corresponding $\textsf{MMD1}(R_l)$. Then it must give an $\alpha$-approximate solution $\textbf{f}_{\alpha}(R_l)$ to $\textsf{MMD1}(R_l)$. Moreover, $\textbf{f}_{\alpha}(R_l)$ must be a feasible periodically repeated solution to $\textsf{MPA}(R_l,R_u,D)$ and $\textsf{MAA}(R_l,R_u,D)$, with an approximation ratio of $(\alpha+\mathsf{c})$ where $\mathsf{c}$ is defined below 
	\be\label{eqn:gamma-beta}
	\mathsf{c}= \begin{cases}
		2\cdot R_u/R_l, & \text{for $\textsf{MPA}(R_l,R_u,D)$}, \\
		3\cdot R_u/R_l, & \text{for $\textsf{MAA}(R_l,R_u,D)$}.
	\end{cases}
	\ee
\end{theorem}
\end{mdframed}
\begin{proof}
	Refer to Appendix~\ref{subsec:MLP-AOI}.
\end{proof}

Thm.~\ref{thm:MLP-AOI} shows that for any $\alpha$-approximation algorithm of the min-max-delay flow problem, we can directly use it to solve \textsf{MPA} and \textsf{MAA} approximately instead, with approximation ratios determined by $\alpha$, $R_{l}$, and $R_{u}$. Note that approximation algorithms for the min-max-delay flow problem exist in the literature, e.g., Misra \emph{et al.}~\cite{misra2009polynomial} have designed a $(1+\epsilon)$-approximation algorithm, where $\epsilon$ can be an arbitrary user-defined positive number.
\section{Performance Evaluation}\label{sec:experiments}
\begin{figure}[]
	\centering
	\vspace{-2ex}
	\includegraphics[width=0.8\linewidth]{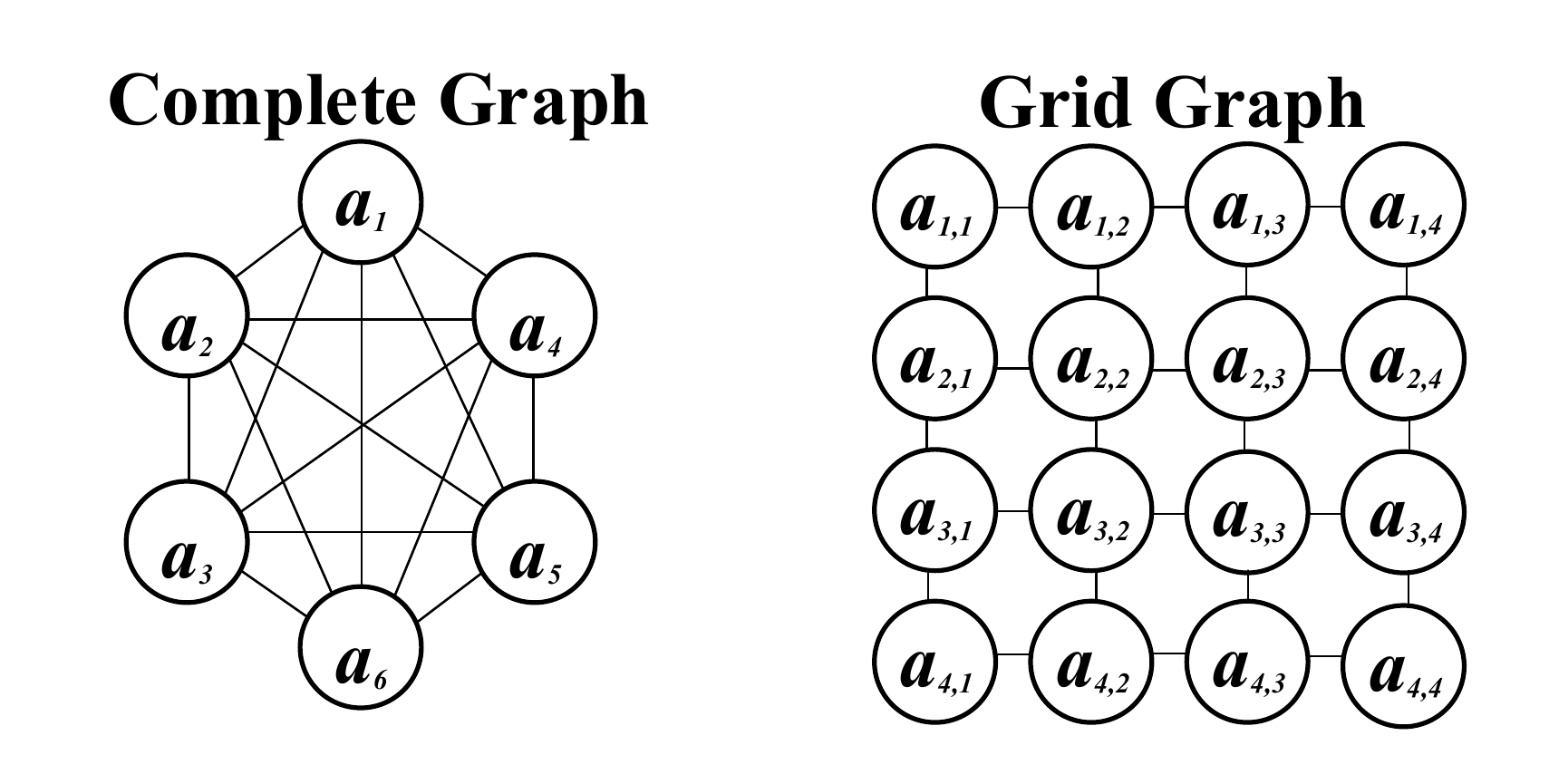}
	\vspace{-2ex}
	\caption{Two simulated network topologies.}
	\label{fig:network}
\end{figure}
We evaluate the empirical performance of our proposed approaches, by simulating (i) two typical network topologies shown in Fig.~\ref{fig:network}, and (ii) nine random network topologies generated by well-known random graph generation models. All networks are modeled as undirected graphs, where each undirected link is treated as two directed links that operate independently. Each link delay is randomly generated from $\{1,2,3,4,5\}$, and each link bandwidth is randomly generated from $\{10,20,30,40,50\}$. Given a network, we consider two different $D$ with $D_1=5\cdot\mathcal{D}$ and $D_2=10\cdot\mathcal{D}$, where $\mathcal{D}$ is the maximum amount of data that can be streamed from sender to receiver with a unit task activation period. Note that this $\mathcal{D}$ is also the maximum throughput that can be achieved in each simulation, based on Lem.~\ref{lem:MLP}. Thus $5$ (resp. $10$) is the minimal possible task activation period for simulations with $D=D_1$ (resp. with $D=D_2$). In each simulation, we consider ten different task activation periods (thus ten different throughputs), where $\mathcal{T}(f)\in\{5,6,...,14\}$ (resp. $\mathcal{T}(f)\in\{10,11,...,19\}$) for simulations with $D=D_1$ (resp. with $D=D_2$). 
The $\textsf{ALG-MMD1}(\alpha)$ used by our approximation framework is the $(1+\epsilon)$-approximation algorithm from~\cite{misra2009polynomial} and we set $\epsilon=1$. Our test environment is an Intel Core i5 (2.40 GHz) processor with 8 GB memory. All the experiments are implemented in C++ and linear programs are solved using \emph{CPLEX}~\cite{cplex}.

\subsection{Simulations on Typical Networks}
 
\begin{figure}[]
	\centering
	\vspace{-2ex}
	\subfigure[Complete graph with $D=D_1$. \label{subfig:com-5}]
	{\includegraphics[width=0.45\linewidth]{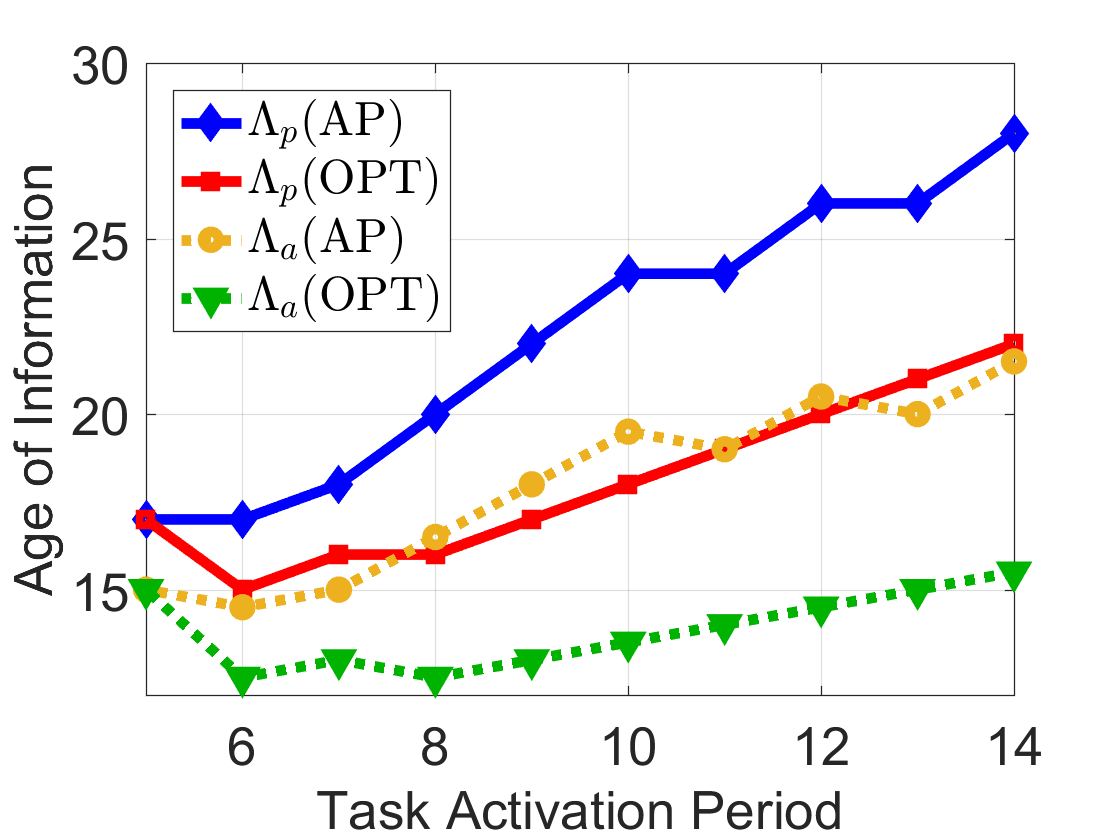}}
	\hfill
	\subfigure[Grid graph with $D=D_2$.  \label{subfig:grid-10}]
	{\includegraphics[width=0.45\linewidth]{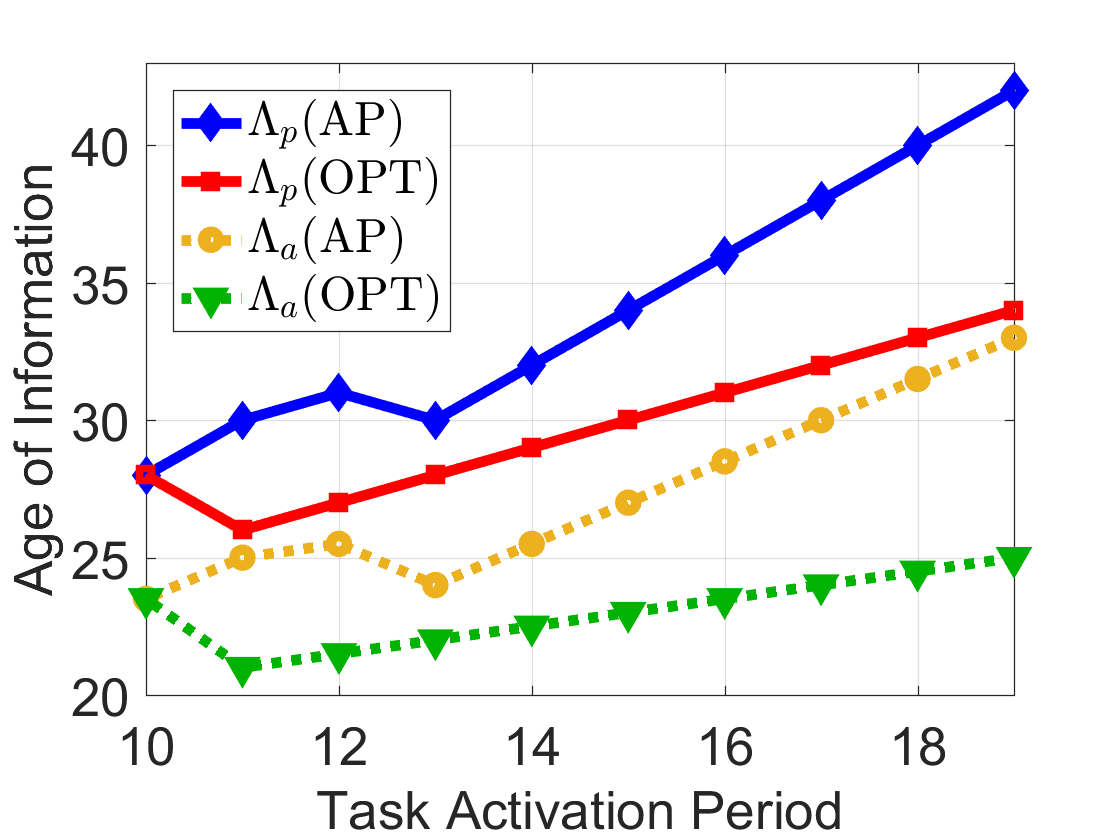}}
	\vspace{-2ex}
	\caption{Simulation results of two representative instances on the typical graphs. 
	}
	\label{fig:typical-graph}
\end{figure}
The two typical network topologies simulated are (i) a complete graph with $6$ nodes and $15$ undirected links, and (ii) a grid graph with $16$ nodes and $24$ undirected links. The complete graph topology represents a fully-connected and thus ideal network structure, while the grid graph topology represents a distributed network structure. In Fig.~\ref{fig:network}, for the complete network, we assume the sender to be $a_1$ and the receiver to be $a_6$, and for the grid network, we assume the sender to be $a_{1,1}$ and the receiver to be $a_{4,4}$.

First, we give the \textsf{AoI} results of one representative instance simulated on the complete graph with $D=D_1$ (resp. on the grid graph with $D=D_2$) in Fig.~\ref{subfig:com-5} (resp. Fig.~\ref{subfig:grid-10}), where for each throughput $R$ (thus for each task activation period $D/R$), $\Lambda_p(\textsf{AP})$ (resp. $\Lambda_a(\textsf{AP})$) is the peak \textsf{AoI} (resp. average \textsf{AoI}) of the solution of $\textsf{ALG-MMD1}(2)$ with a throughput requirement of $R$, while $\Lambda_p(\textsf{OPT})$ (resp. $\Lambda_a(\textsf{OPT})$) is exactly $\Lambda_{p}^{R}$ (resp. $\Lambda_{a}^{R}$), which is the peak \textsf{AoI} (resp. average \textsf{AoI}) of the solution of our Algorithm~\ref{alg:pseudo}.

From Fig.~\ref{subfig:com-5}, empirically we verify (i) Lem.~\ref{lem:peak-average}, where the task activation period of $6$ (thus the throughput of $D_1/6$) achieving the optimal peak \textsf{AoI} is different from that of $8$ (resp. that of $D_1/8$) achieving the optimal average \textsf{AoI}, and (ii) Lem.~\ref{thm:age-T}, where the minimal peak \textsf{AoI} (resp. minimal average \textsf{AoI}) is non-monotonic, non-convex, and non-concave with throughput.

\begin{figure}[]
	\centering
	\vspace{-2ex}
	\subfigure[Complete graph with $D=D_2$. \label{subfig:general-com-10}]
	{\includegraphics[width=0.45\linewidth]{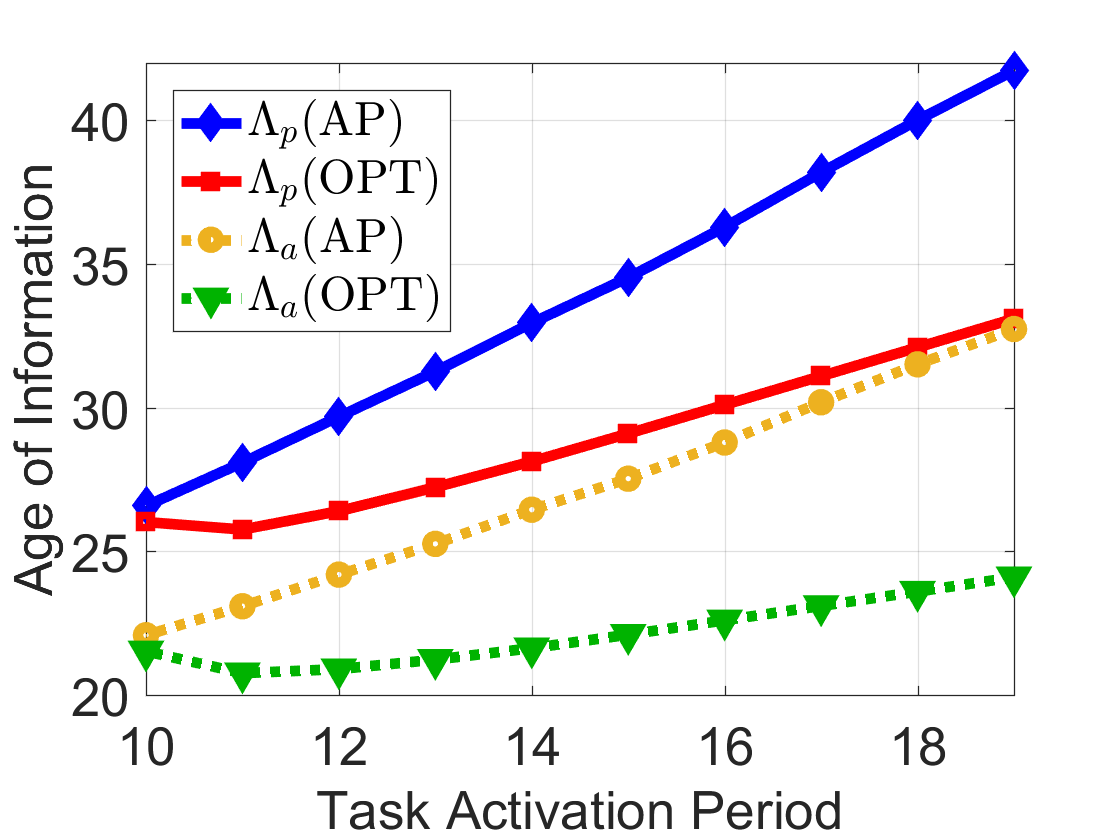}}
	\hfill
	\subfigure[Grid graph with $D=D_1$.  \label{subfig:general-grid-5}]
	{\includegraphics[width=0.45\linewidth]{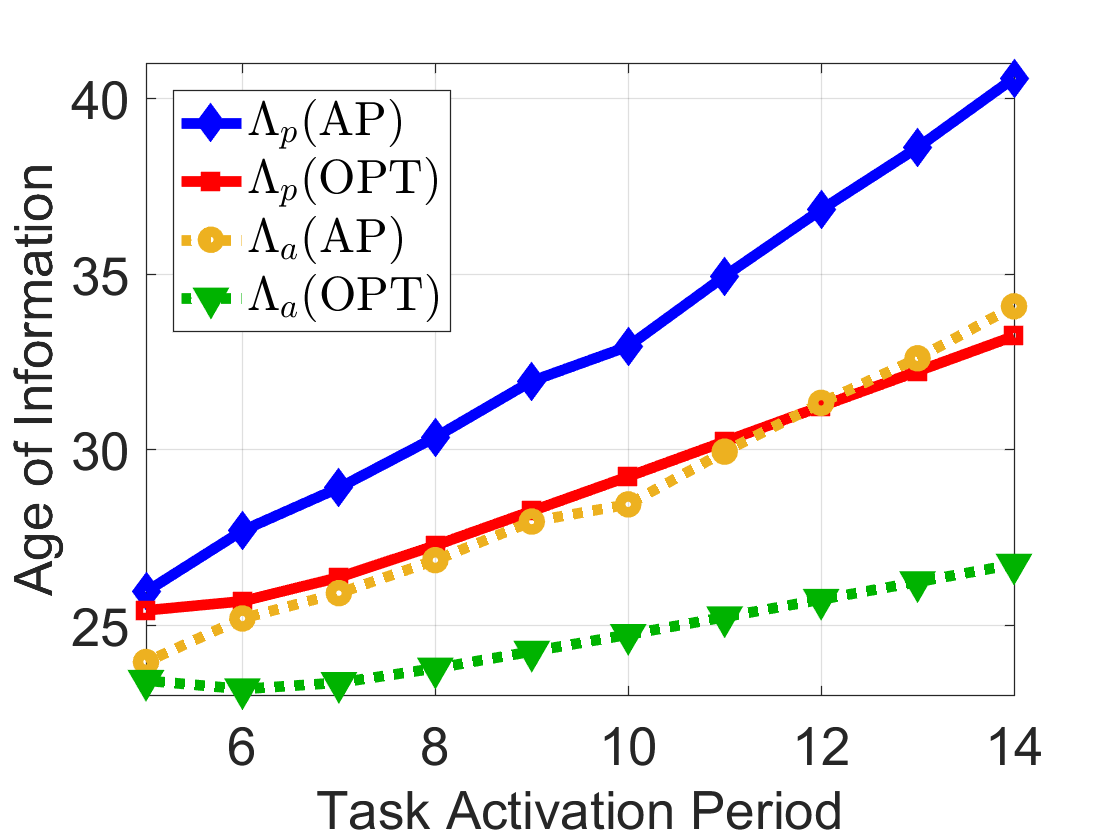}}
	\vspace{-2ex}
	\caption{Simulation results in average of $200$ instances on the typical graphs.
	}
	\label{fig:general}
\end{figure}

Considering that we generate link bandwidths and delays randomly, next, we simulate $100$ instances of the complete network with $D=D_2$ (resp. $100$ instances of the grid network with $D=D_1$), and present the \textsf{AoI} results in average in Fig.~\ref{subfig:general-com-10} (resp. Fig.~\ref{subfig:general-grid-5}). (i) Empirically, we observe that $\Lambda_p^R$ and $\Lambda_a^R$ are ``almost" increasing with throughput $R$. Note that for an instance of \textsf{MPA} (resp. \textsf{MAA}), the peak \textsf{AoI} (resp. average \textsf{AoI}) of our approximation framework is the $\Lambda_p(\textsf{AP})$ (resp. $\Lambda_a(\textsf{AP})$) corresponding to the smallest throughput (thus the largest task activation period), while the peak \textsf{AoI} (resp. average \textsf{AoI}) of our optimal algorithm is the smallest peak \textsf{AoI} (resp. average \textsf{AoI}) among those achieved by all possible throughputs (thus all possible task activation periods). (ii) Empirically, we observe that our optimal algorithm obtains a $3.8\%$ peak \textsf{AoI} reduction (resp. $3.2\%$ average \textsf{AoI} reduction) as compared to our approximation framework, when the number of possible throughputs (thus the range of task activation period) of an instance of \textsf{MPA} (resp. \textsf{MAA}) increases by $1$. However, (iii) given a specific throughput, the average running time of $\textsf{ALG-MMD1}(2)$ (resp. of Algorithm~\ref{alg:pseudo}) is $0.06$s (resp. $0.10$s). Therefore for an instance of \textsf{MPA} and \textsf{MAA}, the running time of our approximation framework is a constant $0.06$s (directly run $\textsf{ALG-MMD1}(2)$ with the smallest throughput requirement), while that of our optimal algorithm increases by $0.10$s when the number of possible throughputs increases by $1$ (enumerate \textsf{AoI}s achieved by all possible throughputs to figure out the optimal).    

\subsection{Simulations on Random Networks}
We also use \emph{SNAP}~\cite{snap} to randomly generate nine network topologies, where three of them follow Erdos-Renyi model, another three of them follow Watts-Strogatz model, and the remaining three of them follow Copying model. For the model-related parameters, we set $n=20$ which is the number of nodes and $m=50$ which is the number of (undirected) links. Besides, we use default values both of the degree parameter $k=3$ and of the degree-exponent parameter $p=0.1$. Definitions of those graph generation models and associated parameters are given by~\cite{snap}.

\begin{figure}[]
	\centering
	\vspace{-2ex}
	\subfigure[Instances with $D=D_1$. \label{subfig:random-5}]
	{\includegraphics[width=0.45\linewidth]{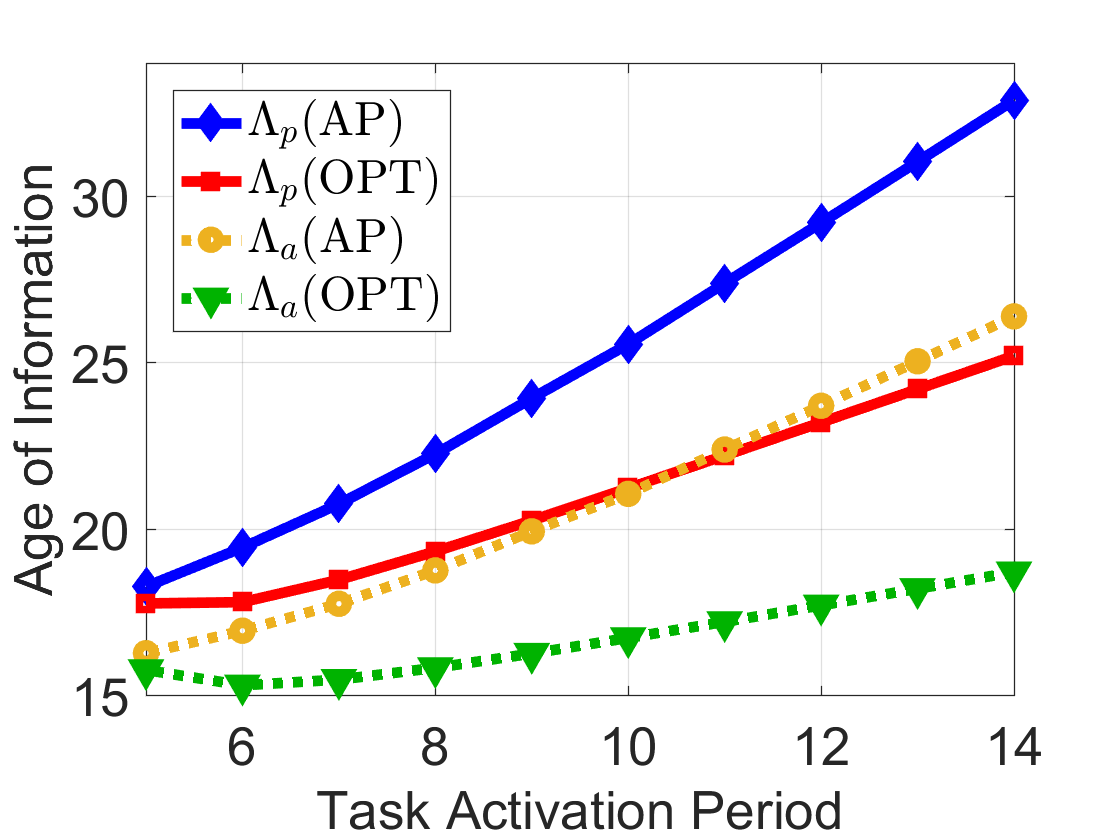}}
	\hfill
	\subfigure[Instances with $D=D_2$.  \label{subfig:random-10}]
	{\includegraphics[width=0.45\linewidth]{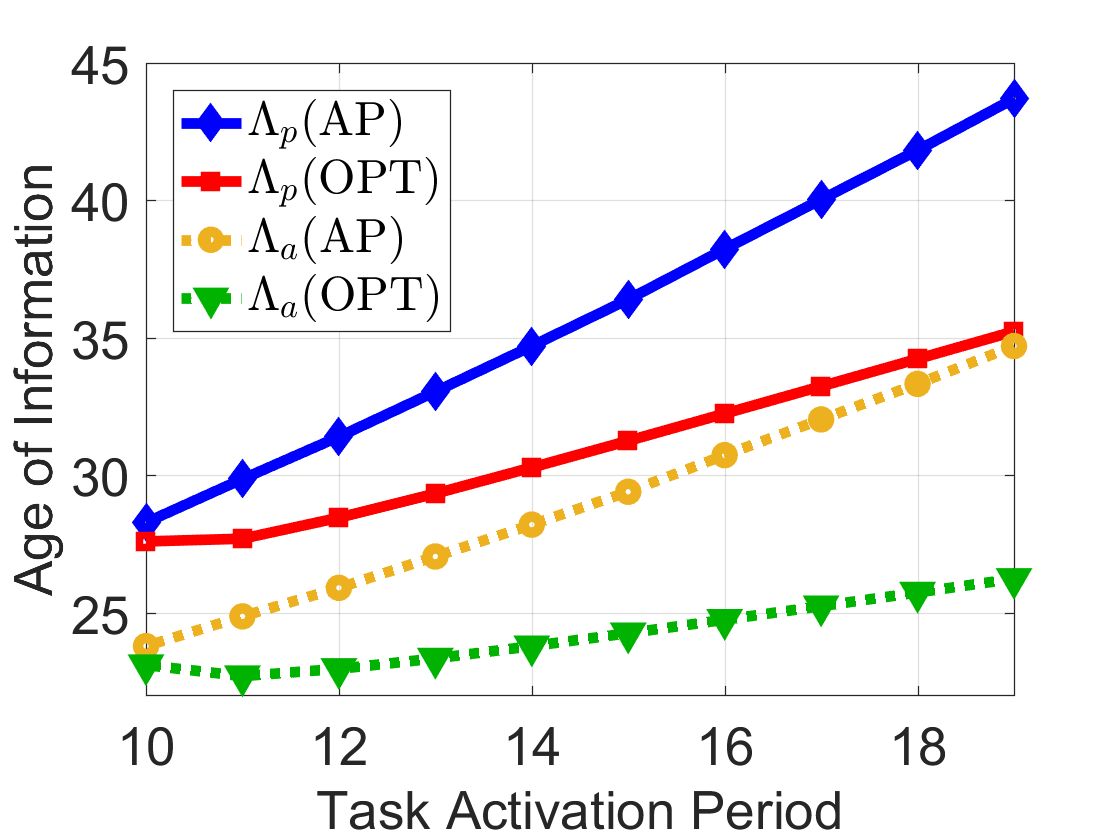}}
	\vspace{-2ex}
	\caption{Simulation results in average of random graphs that are generated by \emph{SNAP}~\cite{snap}. 
	}
	\label{fig:random}
\end{figure}

For each of the nine topologies, we run $100$ simulation instances respectively with $D=D_1$ and with $D=D_2$. Note that for each simulation instance, the sender and the receiver are randomly selected. The simulated \textsf{AoI} results on random networks (Fig.~\ref{fig:random}) is very similar to that on typical networks (Fig.~\ref{fig:general}). When the range of task activation period of an instance increases by $1$, (i) our optimal algorithm obtains a $4.3\%$ peak \textsf{AoI} reduction (resp. $4.0\%$ average \textsf{AoI} reduction) as compared to our approximation framework; (ii) our approximation framework has a constant running time of $0.06$s, while the running time of our optimal algorithm increases by $0.11$s. 

\section{Conclusion}\label{sec:conclusion}
We consider a scenario where a sender periodically sends a batch of data to a receiver over a multi-hop network using multiple paths. We study problems of minimizing peak/average \textsf{AoI}, by jointly optimizing (i) the throughput subject to throughput requirements, and (ii) the multi-path routing strategy. The consideration of batch generation and multi-path communication differentiates our study from existing ones. First we show that our problems are NP-hard but only in the weak sense, as we develop a pseudo-polynomial-time optimal algorithm. Next, we show that minimizing \textsf{AoI} is ``largely" equivalent to minimizing maximum delay, as any optimal solution of the latter is an approximate solution to the former, with bounded optimality loss. We leverage this understanding to design a framework to adapt any polynomial-time $\alpha$-approximation algorithm of the maximum delay minimization problem to solve our \textsf{AoI} minimization problems, with an approximation ratio of $\alpha+\mathsf{c}$. The framework suggests a new avenue for developing approximation algorithms for minimizing \textsf{AoI} in multi-path communications. We conduct extensive simulations over various network topologies to empirically validate the effectiveness of our approach. 

\bibliographystyle{ACM-Reference-Format}
\bibliography{References}
\newpage
\section{Appendix}
\subsection{Different Kinds of Networking Delay in the Slotted Data Transmission Model}\label{subsec:delay}
Slotted data transmission model can take different kinds of networking delay into consideration, due to the following concerns which have been discussed in another study~\cite{bai2012dear}.

(1) Link propagation delay. It is the time for a signal to pass the link, and can be computed as the ratio between the link period and the propagation speed. Clearly that it is a non-negative constant for each link $e\in E$ and can be counted into our link delay $d_e$.

(2) Link transmission delay. It is the time required to push the data onto the link. In general, the link transmission delay is given by $S_e/b_e$ where $S_e$ is the size of the data that is required to be pushed onto the link $e=(v,w)\in E$ at time $t$. In our slotted transmission model, if $S_e\le b_e$, the incurred link transmission delay is $\lceil S_e/b_e\rceil=1$ and is counted into $d_e$. Otherwise if $b_e<S_e\le 2b_e$, the link bandwidth constraint restricts that $e$ first pushes $b_e$ amount of data onto it at time $t$, experiencing a transmission delay of $1$ that is counted into $d_e$. Then the remaining $S_e-b_e$ amount of data will be held at the node $v$ for one time slot, and is pushed to $e$ at time $t+1$, experiencing a transmission delay of $2$, where $1$ delay is counted into $d_e$, and the remaining $1$ delay is incurred by holding the data at node $v$ from time $t$ to time $t+1$. Similarly, we can figure out the transmission delay for arbitrary $S_e: (k-1)b_e<S_e\le kb_e,\forall k\in\mathbb{Z}^+$.

(3) Node queuing delay. It is the time the data spends in the routing queue of a node. When data arrives at a router (node), it has to be processed and then transmitted. But due to the finite service rate $\lambda_v$ of a router $v\in V$, the router puts the data into the queue until it can get around to transmitting them if the data arrival rate is larger than $\lambda_v$. If queuing delay is involved in our problem for each router node $v\in V$, we can add a new node $v'$ to $V$ and change all the outgoing links of $v$ to be the outgoing links of $v'$. We also add a new link $(v,v')$ to $E$, set its bandwidth to be $\lambda_v$ and delay to be $1$. Then the queuing delay can be taken into consideration using the updated network, due to similar reasons in our aforementioned discussions on the link transmission delay.
\subsection{Proof to Lem.~\ref{lem:boundedU}}\label{subsec:boundedU}
\begin{proof}	
	Let us denote $x^p(\vec{u})(g)$ as the value of $x^p(\vec{u})$ in $g$, given a path $p$ and the corresponding offset $\vec{u}$. Suppose there exists a $x^{\bar{p}}(\vec{\bar{u}})>0$ in $g$ with $k\cdot \mathcal{T}(g)\le \bar{u}_i-d_{e_{i-1}}-\bar{u}_{i-1}<(k+1)\cdot \mathcal{T}(g),k\in\mathbb{Z}^+$ for certain $p=\bar{p}$, $\vec{u}=\vec{\bar{u}}$, and $i\in\{1,2,...,|p|\}$. We prove this lemma by directly constructing another periodically repeated solution $f$ based on $g$, and show that $f$ meets all results of this lemma. 
	
	First, we let $\mathcal{T}(f)=\mathcal{T}(g)$. Then we set $x^p(\vec{u})(f)=0$ for all $p$ and all $\vec{u}$. Then for each positive $x^p(\vec{u})(g)$, (i) if $\langle p,\vec{u} \rangle\neq \langle \bar{p},\vec{\bar{u}} \rangle$, we let $x^p(\vec{u})(f)=x^p(\vec{u})(f)+x^p(\vec{u})(g)$. (ii) If $\langle p,\vec{u} \rangle= \langle \bar{p},\vec{\bar{u}} \rangle$, we let $x^{\bar{p}}(\vec{\bar{u}}^*)(f)=x^{\bar{p}}(\vec{\bar{u}}^*)(f)+x^{\bar{p}}(\vec{\bar{u}})(g)$, where comparing $\vec{\bar{u}}^*=\{\bar{u}_j^*,j=0,1,...,|\bar{p}|\}$ with $\vec{\bar{u}}=\{\bar{u}_j,j=0,1,...,|\bar{p}|\}$, we have $\bar{u}_j^*=\bar{u}_j$ for $j=0,1,...,i-1$, but $\bar{u}_j^*=\bar{u}_j-k\cdot\mathcal{T}(g)$ for $j=i,i+1,...,|\bar{p}|$.  
	
	First, we prove $f$ is a feasible periodically repeated solution. It is clear that $f$ meets throughput requirements, because $g$ meets throughput requirements, and $\mathcal{T}(f)=\mathcal{T}(g)$. The satisfied link bandwidth constraints of $f$ comes from the fact that for any link $e\in \bar{p}$, if $x^{\bar{p}}(\vec{\bar{u}})(g)$ respects the link bandwidth constraint of $e$ at certain offset $j\in\{0,1,...,\mathcal{T}(g)-1\}$, then $x^{\bar{p}}(\vec{\bar{u}}^*)(f)$ shall respect the link bandwidth constraint of $e$ at the same offset $j$. This is because that the difference between $\bar{u}_j^*$ and $\bar{u}_j$ is a multiple of the task activation period $\mathcal{T}(g)$ ($\mathcal{T}(f)$ equivalently). Thus according to the constraints~\eqref{eqn:capacity}, the satisfied link bandwidth constraints in $g$ implies that link bandwidth constraints in $f$ are met.
	
	Next, recall that by our assumption, $x^{\bar{p}}(\vec{\bar{u}})(g)>0$ but with $k\cdot \mathcal{T}(g)\le \bar{u}_i-d_{e_{i-1}}-\bar{u}_{i-1}<(k+1)\cdot\mathcal{T}(g),k\in\mathbb{Z}^+$ for a specific $i\in\{1,2,...,|\bar{p}|\}$. Now for the constructed $f$, we have (i) for any $\langle p,\vec{u}\rangle\neq \langle\bar{p},\vec{\bar{u}}^*\rangle$ such that $x^{p}(\vec{u})(g)=0$, it holds that $x^{p}(\vec{u})(f)=0$. (ii) We have $x^{\bar{p}}(\vec{\bar{u}})(g)>0$, while $x^{\bar{p}}(\vec{\bar{u}})(f)=0$. And (iii) although $x^{\bar{p}}(\vec{\bar{u}}^*)(f)>0$, it holds that $\bar{u}_i^*-d_{e_{i-1}}-\bar{u}_{i-1}^*\le \mathcal{T}(g)-1$, and $\bar{u}_j^*\le \bar{u}_j$ for any $j=0,1,...,|\bar{p}|$. Those three results prove that (i) $\mathcal{M}(f)\le \mathcal{M}(g)$, implying that $\Lambda_p(f)\le \Lambda_p(g)$ and $\Lambda_a(f)\le \Lambda_a(g)$ according to Lem.~\ref{lem:age-delay}. (ii) In $g$ for certain node, the data-holding is larger than $\mathcal{T}(g)-1$ slots at certain offset, while the data-holding delay is reduced be no more than $\mathcal{T}(f)-1$ slots for the same node at the same offset in the solution $f$.	
\end{proof}
\subsection{Proof to Lem.~\ref{lem:age-delay}}\label{subsec:age-delay}
\begin{proof}
	Following a periodically repeated solution $f$, the data of the period starting at time $k\cdot\mathcal{T}(f)$ must be received by the receiver at time $k\cdot\mathcal{T}(f)+\mathcal{M}(f)$, and the data of the next period that starts at time $(k+1)\cdot\mathcal{T}(f)$ must be received by the receiver at time $(k+1)\cdot\mathcal{T}(f)+\mathcal{M}(f)$. For arbitrary $k\in\mathbb{Z}$, according to the definition~\eqref{eqn:age}, the \textsf{AoI} at time $k\cdot\mathcal{T}(f)+\mathcal{M}(f)$ and time $(k+1)\cdot\mathcal{T}(f)+\mathcal{M}(f)$ both are $\mathcal{M}(f)$, while it increases linearly from $\mathcal{M}(f)$ to $\mathcal{M}(f)+\mathcal{T}(f)-1$ from the time $k\cdot\mathcal{T}(f)+\mathcal{M}(f)$ to the time $(k+1)\cdot\mathcal{T}(f)+\mathcal{M}(f)-1$. Therefore, it is clear that the peak \textsf{AoI} of $f$ is $\mathcal{M}(f)+\mathcal{T}(f)-1$, which appears periodically at time $k\cdot\mathcal{T}(f)+\mathcal{M}(f)-1,k\in\mathbb{Z}$, and the following holds for the average \textsf{AoI} of $f$
	\bee
	\Lambda_a(f) && =\frac{\sum_{t\in\mathbb{Z}}\mathcal{I}(f,t)}{\sum_{t\in\mathbb{Z}}1}=\lim_{k\to +\infty}\frac{k\cdot\sum_{i=0}^{\mathcal{T}(f)-1}(\mathcal{M}(f)+i)}{k\cdot\mathcal{T}(f)}\nnb\\
	&& =\mathcal{M}(f)+\frac{\mathcal{T}(f)-1}{2},\nnb
	\eee
	which completes the proof.
\end{proof}
\subsection{Proof to Lem.~\ref{lem:delay-aoi}}\label{subsec:delay-aoi}
\begin{figure}[]
	\centering
	\includegraphics[width=0.45\columnwidth]{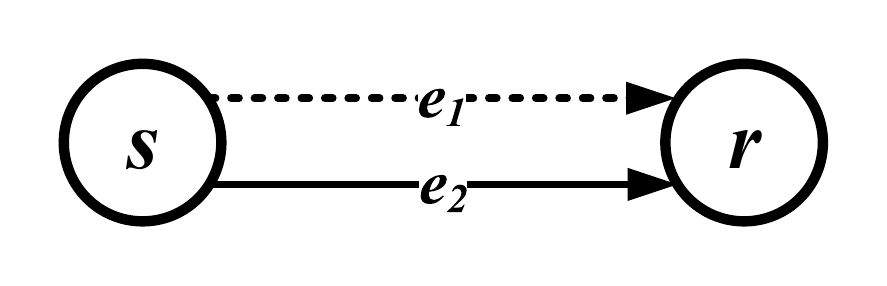}
	\caption{Assume the delay of $e_1$ (resp. $e_2$) is 1 (resp. 11), the bandwidth of $e_1$ (resp. $e_2$) is 1 (resp. 10), $D=10$, $R_l=1$, and $R_u=10/7$.}
	\label{fig:delay-aoi}
\end{figure}
\begin{proof}
	\textbf{First} we prove the result 1 by contradiction. 
	
	Let us assume that $R_p^*<R_m^*$ for certain $R_p^*\in \vec{R}_p$ and $R_m^*\in \vec{R}_m$. Because $R_m^*$ minimize maximum delay, we have
	\be
	\mathcal{M}^{R_m^*}~\le~ \mathcal{M}^{R_p^*}.\label{eqn:use1-1-1}
	\ee
	
	Because $R_p^*$ minimizes peak \textsf{AoI}, according to Lem.~\ref{lem:age-delay}, we have
	\be
	\mathcal{M}^{R_p^*}+D/R_p^*-1~\le~ \mathcal{M}^{R_m^*}+D/R_m^*-1,\nnb
	\ee
	further implying that
	\be
	\mathcal{M}^{R_p^*}-\mathcal{M}^{R_m^*}~\le~ D/R_m^*-D/R_p^*~\overset{(a)}{< }~0,\label{eqn:use1-1-2}
	\ee
	where the inequality (a) comes from our assumption that $R_p^*<R_m^*$. We observe that the inequality~\eqref{eqn:use1-1-1} contradicts with the inequality~\eqref{eqn:use1-1-2}. Thus we must have
	\be
	\min_{R_p\in \vec{R}_p}R_p~\ge~ \max_{R_m\in \vec{R}_m}R_m.\nnb
	\ee
	
	Following the same method, we can prove the following
	\be
	\min_{R_a\in \vec{R}_a}R_a~\ge~ \max_{R_m\in \vec{R}_m}R_m.\nnb
	\ee
	
	\textbf{Next} we prove the result 2 using the example in Fig.~\ref{fig:delay-aoi} with $R_l=1$ and $R_u=10/7$, leading to a maximum task activation period of $10$ and a minimum task activation period of $7$. 
	
	Given a throughput of $R=1$ and thus a task activation of $T=10$, by streaming $1$ data to the link $e_1$ at offsets $i:i=0,1,...,9$, we have $\mathcal{M}^{1}=10$, $\Lambda_{p}^1=19$, and $\Lambda_{a}^1=14.5$. Given $R=10/9$ and thus $T=9$, by streaming $1$ data to the link $e_1$ at offsets $i:i=0,1,...,8$, and streaming $1$ data to the link $e_2$ at the offset $0$, we have $\mathcal{M}^{10/9}=11$, $\Lambda_{p}^{10/9}=19$, and $\Lambda_{a}^{10/9}=15$. Similarly, we have $\mathcal{M}^{10/8}=11$, $\Lambda_{p}^{10/8}=18$, and $\Lambda_{a}^{10/8}=14.5$. And we have $\mathcal{M}^{10/7}=11$, $\Lambda_{p}^{10/7}=17$, and $\Lambda_{a}^{10/7}=14$. Overall, in this instance it is clear that $\vec{R}_p=\vec{R}_a=\{10/7\}$, different from $\vec{R}_m=\{1\}$.  
\end{proof}
\subsection{Proof to Lem.~\ref{lem:delay-aoi-gap}}\label{subsec:delay-aoi-gap}
\begin{proof}
	\textbf{First}, for any $R_m\in\vec{R}_m$ and $R_p\in\vec{R}_p$, we have
	\bee
	\Lambda_{p}^{R_m}-\Lambda_{p}^{R_p} && =\mathcal{M}^{R_m}+D/R_m-1-\left(\mathcal{M}^{R_p}+D/R_p-1\right)\nnb\\
	&& =\left(\mathcal{M}^{R_m}-\mathcal{M}^{R_p}\right)+D/R_m-D/R_p\nnb\\ && \overset{(a)}{\le}D/R_m-D/R_p \le D/R_{l}-D/R_{u},\nnb
	\eee
	which proves the gap~\eqref{eqn:delay-peak-gap}. Here the inequality (a) comes from that $R_m$ minimizes maximum delay.
	
	Following the same method, we can prove the gap~\eqref{eqn:delay-avg-gap}.
	
	\textbf{Second}, for the gap~\eqref{eqn:delay-peak-gap}, it is clear that it is near-tight when $D/R_{l}-D/R_{u}\le 1$. Thus we only focus on cases where $D/R_{l}-D/R_{u}\ge 2$.
	
	Given any $D>0$, $D/R_l\in\mathbb{Z}^+$, and $D/R_u\in\mathbb{Z}^+$, we consider a network with the same topology as Fig.~\ref{fig:delay-aoi}, but assume the delay of $e_1$ (resp. $e_2$) is 1 (resp. $D/R_{l}+1$), the bandwidth of $e_1$ (resp. $e_2$) is 1 (resp. $D/R_{l}$), the amount of batch of data to be sent periodically is $D/R_{l}$, the minimum throughput requirement is $1$, and the maximum throughput requirement is $R_u/R_l$. By streaming 1 data to $e_1$ at offsets $i:i=0,1,...,D/R_{l}-1$, we have $\mathcal{M}^{1}=D/R_{l}$ and $\Lambda_{p}^{1}=2D/R_{l}-1$. By streaming $D/R_{l}$ data to $e_2$ at the offset $0$, we have $\mathcal{M}^{D/(R_l\cdot T)}=D/R_{l}+1$ and $\Lambda_{p}^{D/(R_l\cdot T)}=D/R_{l}+T$, for any task activation period $T\in[D/R_u,D/R_l-1],T\in\mathbb{Z}^+$. It is clear that in this example, $\vec{R}_m=\{1\}$ and $\vec{R}_p=\{R_u/R_l\}$. 
	\be
	\Lambda_{p}^{R_m}-\Lambda_{p}^{R_p}=2\cdot \frac{D}{R_l}-1-\left(\frac{D}{R_l}+\frac{D}{R_u}\right)\ge \frac{D}{R_l}-\frac{D}{R_u}-1.\nnb
	\ee
	
	The gap~\eqref{eqn:delay-avg-gap} is obviously near-tight for cases where $D/R_{l}-D/R_{u}\le 2$. And the near-tightness of the gap~\eqref{eqn:delay-avg-gap} for cases when $D/R_{l}-D/R_{u}\ge 3$ can be proved using the same instance. 
\end{proof}
\subsection{Proof to Lem.~\ref{lem:peak-average}}\label{subsec:peak-average}
\begin{figure}[]
	\centering
	\includegraphics[width=0.45\columnwidth]{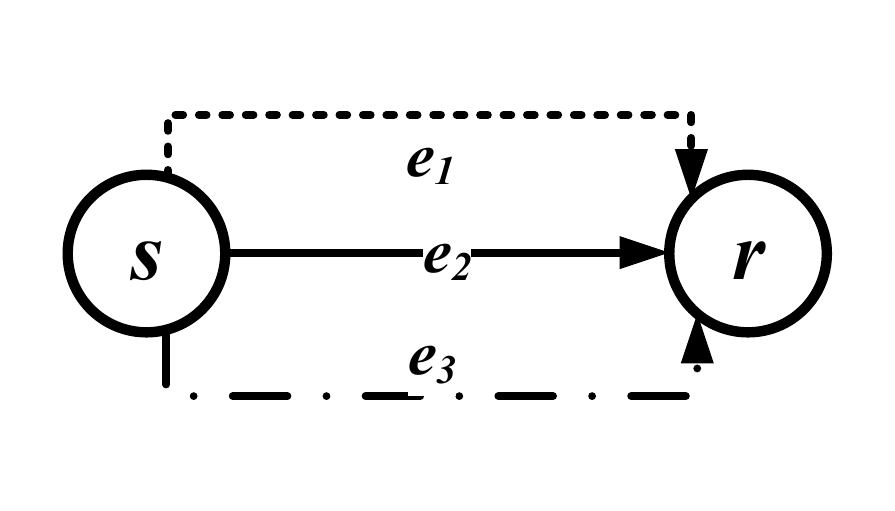}
	\caption{Assume the bandwidth of each of the three links is $1$. Assume the delay of $e_1$, $e_2$, and $e_3$ are $1$, $6$, and $7$, respectively. And assume $D=5$, $R_l=1$, and $R_u=5/2$.}
	\label{fig:peak-average}
\end{figure}
\begin{proof}
	\textbf{First} we prove $R_p\ge R_a$ and $\mathcal{M}^{R_p}\ge 
	\mathcal{M}^{R_a}$ by contradiction. Let us assume that $R_p^*<R_a^*$ for certain $R_p^*\in \vec{R}_p$ and $R_a^*\in \vec{R}_a$. Because $R_p^*$ minimizes peak \textsf{AoI}, according to Lem.~\ref{lem:age-delay}, we have
	\be
	\mathcal{M}^{R_p^*}+D/R_p^*-1~\le~ \mathcal{M}^{R_a^*}+D/R_a^*-1.\label{eqn:use1}
	\ee
	
	Similarly, because $R_a^*$ minimizes average \textsf{AoI}, we have
	\be
	\mathcal{M}^{R_p^*}+\frac{D/R_p^*-1}{2}~\ge~ \mathcal{M}^{R_a^*}+\frac{D/R_a^*-1}{2}.\label{eqn:use2}
	\ee
	
	The above inequality~\eqref{eqn:use1} implies that
	\be
	D/R_a^*-D/R_p^*~\ge~\mathcal{M}^{R_p^*}-\mathcal{M}^{R_a^*}.\label{eqn:use3}
	\ee
	
	Now let us look at the following inequality
	\bee
	&& \mathcal{M}^{R_p^*}+\frac{D/R_p^*-1}{2}-\left(\mathcal{M}^{R_a^*}+\frac{D/R_a^*-1}{2}\right) =\left(\mathcal{M}^{R_p^*}-\mathcal{M}^{R_a^*}\right)\nnb\\
	&& -\frac{D/R_a^*-D/R_p^*}{2} \overset{(a)}{\le}\frac{D/R_a^*-D/R_p^*}{2}\overset{(b)}{<}0,\label{eqn:use4}
	\eee
	where inequality (a) holds due to inequality~\eqref{eqn:use3}, and the inequality (b) comes from our assumption that $R_p^*<R_a^*$. It is clear that the inequality~\eqref{eqn:use2} is contradicted with the inequality~\eqref{eqn:use4}, implying that it must hold that $R_p\ge R_a$ for all $R_p\in \vec{R}_p$ and $R_a\in \vec{R}_a$.
	
    For maximum delay, for any $R_p\in \vec{R}_p$ and $R_a\in \vec{R}_a$, we have
    \bee
    \mathcal{M}^{R_a}-\mathcal{M}^{R_p}&& =\Lambda_a^{R_a}-(D/R_a-1)/2-\left(\Lambda_a^{R_p}-(D/R_p-1)/2\right)\nnb\\
    && =\left(\Lambda_a^{R_a}-\Lambda_a^{R_p}\right)+(D/R_p-D/R_a)/2\overset{(a)}{\le}0,\nnb
    \eee
    where inequality (a) holds because that (i) $R_a$ minimizes average \textsf{AoI}, and (ii) $R_a\le R_p$ is true as proved previously.
	
	\textbf{Second}, we prove the gap~\eqref{eqn:gap-1} and~\eqref{eqn:gap-2}. Because any $R_p\in\vec{R}_p$ minimizes peak \textsf{AoI}, we have
	\be
	\mathcal{M}^{R_p}+D/R_p-1\le \mathcal{M}^{R_a}+D/R_a-1,\forall R_p\in\vec{R}_p,\forall R_a\in \vec{R}_a,\nnb 
	\ee
	implying that
	\be
	D/R_a-D/R_p\ge\mathcal{M}^{R_p}-\mathcal{M}^{R_a},\forall R_p\in\vec{R}_p,\forall R_a\in \vec{R}_a.\label{eqn:use1-1} 
	\ee
	
	Therefore, for any $R_p\in\vec{R}_p$ and $R_a\in\vec{R}_a$, we have the following
	\bee
	\Lambda_{a}^{R_p}-\Lambda_{a}^{R_a} && =\mathcal{M}^{R_p}+\frac{D/R_p-1}{2}-\left(\mathcal{M}^{R_a}+\frac{D/R_a-1}{2}\right)\nnb\\
	&& =\left(\mathcal{M}^{R_p}-\mathcal{M}^{R_a}\right)+\frac{D/R_p-D/R_a}{2}\nnb\\
	&& \overset{(a)}{\le}\frac{D/R_a-D/R_p}{2} \le \frac{D}{2R_l}-\frac{D}{2R_u},\nnb
	\eee
	where inequality (a) comes from inequality~\eqref{eqn:use1-1}. We can prove the gap~\eqref{eqn:gap-2} following a similar method, together with an observation that $\Lambda_{p}^{R_a}-\Lambda_{p}^{R_p}$ must be an integer due to our Lem.~\ref{lem:age-delay} considering that all link delays and task activation periods are integers in our slotted data transmission model.
	
	\textbf{Third}, we prove the existence of an instance where $R_p\neq R_a$ for any $R_p\in\vec{R}_p$ and $R_a\in\vec{R}_a$, using the example in Fig.~\ref{fig:peak-average}. 
	
	In Fig.~\ref{fig:peak-average}, it is clear that $\mathcal{M}^1=5$, $\Lambda_{p}^1=9$, and $\Lambda_{a}^1=7$, by streaming $1$ data to the link $e_1$ at offsets $i:i=0,1,...,4$. And we have $\mathcal{M}^{5/4}=6$, $\Lambda_{p}^{5/4}=9$, and $\Lambda_{a}^{5/4}=7.5$, by streaming $1$ data to the link $e_1$ at offsets $i:i=0,1,...,3$, and streaming $1$ data to the link $e_2$ at the offset $0$. Similarly, we have $\mathcal{M}^{5/3}=7$, $\Lambda_{p}^{5/3}=9$, $\Lambda_{a}^{5/3}=8$, $\mathcal{M}^{5/2}=7$, $\Lambda_{p}^{5/2}=8$, and $\Lambda_{a}^{5/2}=7.5$. Now in this example, we have $\vec{R}_p=\{5/2\}$ and $\vec{R}_a=\{1\}$, which are different.
\end{proof}
\subsection{Proof to Lem.~\ref{lem:expanded}} \label{subsec:expanded}
\begin{proof}
	\textbf{Only if part}. Suppose there exists a feasible periodically repeated solution $f=\{x^p(\vec{u}):\forall p,\forall \vec{u}\}$ with $\mathcal{T}(f)=D/R$ and $\mathcal{M}(f)\le M$. Based on $f$, we can construct a flow $\mathbf{f}$ in $G_{\textsf{exp}}$ that is a feasible solution to linear program~\eqref{eqn:Expanded}, and the value of $\mathbf{f}$ is no smaller than $D$, as follows. The source of $\mathbf{f}$ is $s_0$, and the sink of $\mathbf{f}$ is $r_M$. Based on each positive $x^p(\vec{u})$ of $f$ where we assume $p=\langle v_1,v_2,...,v_{|p|}\rangle$ and $\vec{u}=\{u_0,u_1,...,u_{|p|}\}$, (i) we assign a flow rate of $x^p(\vec{u})$ to each of the following links, if it belongs to $E_{\textsf{exp}}$,
	\be
	\left(v_i^{u_i},v_{i+1}^{u_i+d_{e_i}}\right),~ i=1,2,...,|p|-1.\nnb
	\ee
	Each of those links ($(v_i^{u_i},v_{i+1}^{u_i+d_{e_i}})$) is an expanded link corresponding to one link $(v_i,v_{i+1})$ that is on the path $p$.
	(ii) We also assign a rate of $x^p(\vec{u})$ to each of the following links, if it belongs to $E_{\textsf{exp}}$,  
	\be
	\left(v_{i+1}^{u_i+d_{e_i}+j},v_{i+1}^{u_i+d_{e_i}+j+1}\right),j=0,...,u_{i+1}-u_i-d_{e_i}-1,i=0,...,|p|-2.\nnb
	\ee
	which are expanded links denoting that certain node shall hold data at certain offset following the solution $x^p(\vec{u})$ of $f$. (iii) Finally, we assign a rate of $x^p(\vec{u})$ to each of the following links
	\be
	\left(v_{i+1}^{u_i+d_{e_i}+j},v_{i+1}^{u_i+d_{e_i}+j+1}\right),j=0,...,M-u_i-d_{e_i}-1,i=|p|-1,\nnb
	\ee
	which guarantees that the sink of the constructed $\textbf{f}$ is $r^M$.
	
	It is easy to verify that $\mathbf{f}$ meets flow conservation constraints from $s_0$ to $r_M$ in $G_{\textsf{exp}}$. Because $f$ meets the throughput requirements~\eqref{eqn:throughput}, clearly that the value of $\mathbf{f}$ is no smaller than $D$. Because $f$ meets the link bandwidth constraints~\eqref{eqn:capacity}, we have $x_e(i)\le b_e,\forall e\in E,\forall i=0,1,...,D/R-1$ in $f$. By the definition of $e(i)$, we know the sum of rates on $e(i)$ is the aggregated data that is streamed to $e$ at the offset $i$, which is exactly $x_e(i)$. Therefore, the satisfied constraints~\eqref{eqn:capacity} of $f$ implies that the sum of rates on $e(i)$ is no more than $b_e$, for all $e\in E$ and $i=0,1,...,D/R-1$. Overall, we observe that $\mathbf{f}$ is a feasible solution to linear program~\eqref{eqn:Expanded}, and the value of $\mathbf{f}$ is no smaller than $D$, implying that (i) the optimal solution of linear program~\eqref{eqn:Expanded} must exist, and (ii) its value must be no smaller than $D$.
	
	\textbf{If part}. It is proved similarly as the only if part. Suppose linear program~\eqref{eqn:Expanded} outputs an optimal flow whose value is no smaller than $D$. Note that such a flow can be defined on edges in $G_{\textsf{exp}}$. After flow decomposition, we can obtain a flow solution $\mathbf{f}$ defined on paths, i.e., we can get a solution $\mathbf{f}=\{x^p,\forall p\in P_{\textsf{exp}}\}$ where $P_{\textsf{exp}}$ is the set of all simple paths from $s^0$ to $r^M$ in $G_{\textsf{exp}}$, and $x^p$ is the rate assigned to $p$. Note that the size of $\mathbf{f}$, i.e., the number of paths $p\in P_{\textsf{exp}}$ that are assigned positive flow rates, is upper bounded by $|E_{\textsf{exp}}|$~\cite{ford1956maximal}. Based on $\mathbf{f}$, we can construct a feasible periodically repeated solution $f$ with $\mathcal{T}(f)=D/R$, and $\mathcal{M}(f)\le M$. 
	
	First note that there may exist certain $x^p>0$ in $\mathbf{f}$ where the path $p$ can include nodes $v_i^n$ and $v_i^m$ that are expanded nodes of a same node $v\in V$, but with $n< m$. However, in this case, we can obtain an equivalent feasible flow to $\mathbf{f}$, by streaming the positive rate $x^p$ that are originally assigned to a certain outgoing link of $v_i^n$ now to links $(v_i^{n+j},v_i^{n+j+1}),j=0,1,...,m-n-1$, because we do not have bandwidth constraints for those links in the linear program~\eqref{eqn:Expanded}. Thus it is clear that we can always obtain a path $p_f\in P$ (a simple path in $G$) corresponding to a path $p\in P_{\textsf{exp}}$ (a simple path in $G_{\textsf{exp}}$) for any $x^p>0$ in $\mathbf{f}$. By setting $x^{p_f}(\vec{u})=x^p$ where $u_0=0$, $u_{|p_f|}=u_{|p_f|-1}+d_{e_{|p_f|-1}}$, and $u_i\in\vec{u},i=1,2,...,|p_f|-1$ to be the $u\in[0,D],u\in\mathbb{Z}$ such that this path $p$ uses certain incoming link of the node $v_i^u$ but it does not use the link $(v_i^{u},v_{i}^{u+1})$, it is easy to verify that this constructed solution $f$ satisfies the throughput requirements and link bandwidth constraints, with $\mathcal{T}(f)=D/R$, and $\mathcal{M}(f)\le M$. Note that (i) if in the constructed solution $f$ for certain $x^{p_f}(\vec{u})$, there exists an $i$ such that $u_i-d_{e_{i-1}}-u_{i-1}\ge D/R$, we can follow Lem.~\ref{lem:boundedU} to construct another feasible solution that is no worse than $f$ such that $u_i-d_{e_{i-1}}-u_{i-1}\le D/R-1$. (ii) The size of $f$ which is constructed from $\mathbf{f}$ is upper bounded by $|E_{\textsf{exp}}|$, due to that the size of $\mathbf{f}$ is upper bounded by $|E_{\textsf{exp}}|$.  
\end{proof}
\subsection{Proof to Lem.~\ref{lem:time-complexity}}\label{subsec:time-complexity}
\begin{proof}
	The time complexity of solving linear program~\eqref{eqn:Expanded} is $O(|E|^3M_U^3\mathcal{L})$, because the number of variables of the linear program~\eqref{eqn:Expanded} is $|E|M_U$, and the time complexity is $O(n^3\mathcal{L})$ for solving linear programs with $n$ to be the number of variables~\cite{ye1991n3l}. As discussed in the proof to Thm.~\ref{lem:expanded}, the size of the periodically repeated solution $f_t$ to our problem by solving the linear program~\eqref{eqn:Expanded} is upper bounded by $|E_{\textsf{exp}}|$ and it holds that $|E_{\textsf{exp}}|\le |E|M_U$. Those observations, together with the binary-search scheme, imply that the time complexity of Algorithm~\ref{alg:pseudo} is $O(|E|^3M_U^3\mathcal{L}\log M_U)$. 
\end{proof}
\subsection{Proof to Lem.~\ref{thm:age-T}}\label{sub:age-T}
\begin{figure}[]
	\centering
	\includegraphics[width=0.45\columnwidth]{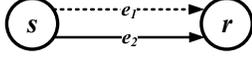}
	\caption{Assume the delay of $e_1$ (resp. $e_2$) is $1$ (resp. $d$), the bandwidth of $e_1$ (resp. $(e_2)$) is $1$ (resp. $5$), $D=5$, $R_l=5/6$, and $R_u=5/3$.}
	\label{fig:age-T}
\end{figure}
\begin{proof}
	We use the example in Fig.~\ref{fig:age-T} to prove this lemma.
	
	\textbf{Non-monotonicity} and \textbf{non-concavity}. Suppose $d=7$. It is clear that $\Lambda_{p}^{1}=9$ and $\Lambda_{p}^{5/6}=10$, by streaming $1$ data to the link $e_1$ at offsets $i:i=0,1,...,4$. Comparing $R_1=1$ with $R_0=5/6$, we see that $\Lambda_{p}^{R}$ is decreasing with $R$. Besides, we have $\Lambda_{p}^{5/4}=10$, by streaming $5$ data to the link $e_2$ at the offset $0$. Comparing $R_2=5/4$ with $R_1=1$, we see that $\Lambda_{p}^R$ is increasing with $R$. Therefore in general, $\Lambda_{p}^R$ is non-monotonic with $R$. Besides, the following holds
	\bee
    \Lambda_{p}^{0.6R_0+0.4R_2} && ~=~\Lambda_{p}^{R_1}~=~9~<~0.6\times 10+0.4\times 10\nnb\\
    && ~=~0.6\cdot \Lambda_{p}^{R_0}+0.4\cdot \Lambda_{p}^{R_2},\nnb
	\eee 
	implying that $\Lambda_{p}^R$ is non-concave with $R$. We can prove the non-monotonicity and non-concavity for $\Lambda_{a}^R$ with $R$ using the same example, because $\Lambda_{a}^{R_2}=8.5$, $\Lambda_{a}^{R_1}=7$, and $\Lambda_{a}^{R_0}=7.5$.
	
	\textbf{Non-convexity}. Suppose $d=6$. We have $\Lambda_{p}^{5/3}=8$, $\Lambda_{p}^{5/4}=9$, and $\Lambda_{p}^1=9$. Considering $R_1=1$, $R_2=5/4$, and $R_3=5/3$, the following holds
	\bee
	\Lambda_{p}^{0.625\cdot R_1+0.375\cdot R_3} && ~=~\Lambda_{p}^{R_2}~=~9~>~0.625\times 9+0.375\times 8\nnb\\
	&& ~=~0.625\cdot \Lambda_{p}^{R_1}+0.375\cdot \Lambda_{p}^{R_3},\nnb
	\eee
	implying that $\Lambda_{p}^R$ is non-convex with $R$. We can prove the non-convexity for $\Lambda_{a}^R$ with $R$ using the same example similarly, considering that $\Lambda_{a}^{R_3}=7$, $\Lambda_{a}^{R_2}=7.5$, and $\Lambda_{a}^{R_1}=7$. 	   
\end{proof}
\subsection{Proof to Lem.~\ref{lem:MLP}}\label{subsec:MLP}
\begin{proof}
	\textbf{First}, we prove the feasibility of $\textsf{MMD1}(R)$. Because $R$ is feasible, there must exist a feasible periodically repeated solution $f=\{x^p(\vec{u}):\forall p\in P,\forall \vec{u}\in\vec{\mathcal{U}}\}$ to $\textsf{MPA}(R_l,R_u,D)$ (resp. $\textsf{MAA}(R_l,R_u,D)$), with $\mathcal{T}(f)=D/R$. We prove the feasibility of $\textsf{MMD1}(R)$ by directly constructing a feasible solution $\textbf{f}=\textbf{f}(R)=\{x^{\textbf{p}}:\forall \textbf{p}\in P\}$ to it based on $f$.
	
	For each $p\in P$, we let $\textbf{p}=p$, and $x^{\textbf{p}}=\sum_{\forall\vec{u}}x^p(\vec{u})\cdot R/D$, i.e., the assigned data on $\textbf{p}$ in $\textbf{f}$ is $(R/D)$-fractional of the aggregated data assigned on the same path $p=\textbf{p}$ over all possible offsets $\vec{u}$ that corresponds to the path $p$ in $f$. Note that for each $x^{\textbf{p}}$ of $\textbf{f}$, we stream $x^{\textbf{p}}$ amount of data over the path $\textbf{p}$ without holding them on any of the nodes that belong to the path $\textbf{p}$.
	
	Because $f$ is feasible to $\textsf{MPA}(R_l,R_u,D)$ (resp. $\textsf{MAA}(R_l,R_u,D)$), we have the following
	\be
	\sum_{\forall p}\sum_{\forall \vec{u}}x^p(\vec{u})~=~D,\nnb
	\ee
	implying that $\sum_{\forall \textbf{p}}x^{\textbf{p}}=R$, i.e., $\textbf{f}$ meets the throughput requirement of $\textsf{MMD1}(R)$ (refer to~\cite{misra2009polynomial} for the detailed definition of the throughput requirement of the min-max-delay flow problem). In order to prove the feasibility of $\textbf{f}$, we only need to prove that it satisfies the link bandwidth constraints of $\textsf{MMD1}(R)$.
	
	Because $f$ is feasible to $\textsf{MPA}(R_l,R_u,D)$ (resp. $\textsf{MAA}(R_l,R_u,D)$), the bandwidth constraints~\eqref{eqn:capacity} must be met, i.e.,
	\be
	\sum_{p:e\in p}\sum_{\substack{k\in\mathbb{Z},\vec{u}:k\cdot \mathcal{T}(f)\\+\mathcal{B}^p(\vec{u},e)=i}} x^p(\vec{u})~\le~b_e,~\forall e\in E,\forall i=0,...,\mathcal{T}(f)-1,\nnb
	\ee
	implying that the following holds
	\be
	\sum_{p:e\in p}\sum_{\vec{u}} x^p(\vec{u})~\le~\mathcal{T}(f)\cdot b_e,~~ \forall e\in E.\nnb
	\ee
	Because for any $p\in P$, we have defined $\textbf{p}=p$ and $x^{\textbf{p}}=\sum_{\vec{u}}x^p(\vec{u})\cdot R/D$, for any $e\in E$, the following shall hold
	\be
	\sum_{\textbf{p}:e\in \textbf{p}}x^{\textbf{p}}=\frac{R}{D}\cdot \sum_{p:e\in p}\sum_{\vec{u}}x^{p}(\vec{u})\le\frac{R}{D}\cdot\mathcal{T}(f)\cdot b_e= b_e,\text{ where }\textbf{p}=p,\nnb
	\ee
	i.e., the bandwidth constraints of $\textsf{MMD1}(R)$ are met (refer to~\cite{misra2009polynomial} for the detailed definition of the link bandwidth constraints of the min-max-delay flow problem). Therefore, $\textbf{f}$ is feasible to $\textsf{MMD1}(R)$, implying that $\textsf{MMD1}(R)$ is theoretically feasible.
	
	\textbf{Second}, suppose $\textbf{f}(R)$ is an arbitrary feasible solution to $\textsf{MMD1}(R)$, it obviously holds that it is a feasible periodically repeated solution to $\textsf{MPA}(R_l,R_u,D)$ (resp. $\textsf{MAA}(R_l,R_u,D)$), achieving a throughput of $R$, since $\textbf{f}(R)$ successfully streams $D$ amount of data from $s$ to $G$ every $D/R$ time slots, meeting link bandwidth constraints. Note that when looking at $\textbf{f}(R)=\{x^{\textbf{p}}:\forall \textbf{p}\in P\}$ which is a solution to $\textsf{MMD1}(R)$ from the perspective of $\textsf{MPA}(R_l,R_u,D)$ (resp. $\textsf{MAA}(R_l,R_u,D)$), although it is not a solution described in the way of formula~\eqref{eqn:solution}, clearly it is periodically repeated, because $\textbf{f}(R)$ requires us that (i) we do not hold any data at any node except for the sender $s$, and (ii) for each path $\textbf{p}\in P$, we stream $x^{\textbf{p}}$ amount of data from $s$ to $\textbf{p}$ at each time slot. From the perspective of $\textsf{MMD1}(R)$, according to the definition of $\hat{\mathcal{M}}(\textbf{f})$~\cite{misra2009polynomial}, we have
	\be
	\hat{\mathcal{M}}(\textbf{f}(R))~=~\max_{\textbf{p}\in P:x^{\textbf{p}}>0}\sum_{e\in E:e\in\textbf{p}}d_e,\nnb
	\ee
	i.e., $\hat{\mathcal{M}}(\textbf{f}(R))$ is the path delay of the slowest path that is assigned a positive flow rate of $\textbf{f}(R)$. But when we look at $\textbf{f}(R)$ from the perspective of $\textsf{MPA}(R_l,R_u,D)$ (resp. $\textsf{MAA}(R_l,R_u,D)$), it is clear that
	\be
	\hat{\mathcal{M}}(\textbf{f}(R))+D/R-1~=~\mathcal{M}(\textbf{f}(R)).\nnb
	\ee
	This is because (i) there is $R$ amount of data generated at the sender for transmission at each time when looking at $\textbf{f}(R)$ as a solution to $\textsf{MMD1}(R)$, while there is $D$ amount of data generated at the sender for transmission every $D/R$ time slots when looking at $\textbf{f}(R)$ as a solution to $\textsf{MPA}(R_l,R_u,D)$ (resp. $\textsf{MAA}(R_l,R_u,D)$). And thus (ii) when we look at $\textbf{f}(R)$ from the perspective of $\textsf{MPA}(R_l,R_u,D)$ (resp. $\textsf{MAA}(R_l,R_u,D)$), it requires us to hold certain positive amount of data at the sender till the offset $D/R-1$, and then use paths with delays upper bounded by $\hat{\mathcal{M}}(\textbf{f}(R))$ to send them from the sender to the receiver. 
\end{proof}
\subsection{Proof to Thm.~\ref{thm:MLP-AOI}} \label{subsec:MLP-AOI}
\begin{proof}
    In this proof, we denote $\textbf{f}^*(R)$ as the optimal solution to $\textsf{MMD1}(R)$.
    
    \textbf{First}, we prove the existence of $\textbf{f}_{\alpha}(R_l)$. According to Lem.~\ref{lem:MLP}, because $R_p$ (resp. $R_a$) is a feasible throughput to $\textsf{MPA}(R_l,R_u,D)$ (resp. $\textsf{MAA}(R_l,R_u,D)$), we know $\textsf{MMD1}(R_p)$ (resp. $\textsf{MMD1}(R_a)$) must be feasible, i.e., $\textbf{f}^*(R_p)$ (resp. $\textbf{f}^*(R_a)$) must exist. Now considering that $R_l\le R_p$ (resp. $R_l\le R_a$), after we delete certain amount of flow rate from the solution $\textbf{f}^*(R_p)$ (resp. $\textbf{f}^*(R_a)$), we can get a feasible solution to $\textsf{MMD1}(R_{l})$, implying that $\textsf{MMD1}(R_{l})$ is theoretically feasible. Therefore, when we use the algorithm $\textsf{ALG-MMD1}(\alpha)$ to solve $\textsf{MMD1}(R_{l})$, we must obtain a feasible solution $\textbf{f}_{\alpha}(R_{l})$.
	
	\textbf{Second}, also based on Lem.~\ref{lem:MLP}, it is obvious that $\textbf{f}_{\alpha}(R_{l})$ is a feasible periodically repeated solution to $\textsf{MPA}(R_l,R_u,D)$ (resp. $\textsf{MAA}(R_l,R_u,D)$), achieving a throughput of $R_l$.

	\textbf{Third}, we prove the approximation ratio $(\alpha+\mathsf{c})$ of $\textbf{f}_{\alpha}(R_l)$ with $\textsf{MPA}(R_l,R_u,D)$. Consider the following inequality
	\be
	\hat{\mathcal{M}}(\textbf{f}_{\alpha}(R_{l})) \overset{(a)}{\le}\alpha\cdot\hat{\mathcal{M}}(\textbf{f}^*(R_{l}))\overset{(b)}{\le}\alpha\cdot\hat{\mathcal{M}}(\textbf{f}^*(R_p)),\label{eqn:epsilon}
	\ee
	where $\hat{\mathcal{M}}(\textbf{f})$ is same defined as in Lem.~\ref{lem:MLP}. Here inequality (a) holds because $\textsf{ALG-MMD1}(\alpha)$ is an $\alpha$-approximation algorithm for $\textsf{MMD1}(R_l)$. Inequality (b) is true because the minimal maximum delay of the min-max-delay flow problem is non-decreasing with the input throughput requirement, considering that given $R_1\ge R_2$, after deleting certain amount of flow rate from $\textbf{f}^*(R_1)$ that is optimal to $\textsf{MMD1}(R_1)$, we can obtain a feasible solution to $\textsf{MMD1}(R_2)$, whose maximum delay is obviously no smaller than $\textbf{f}^*(R_2)$.

	Suppose $f_p$ is the optimal solution to $\textsf{MPA}(R_l,R_u,D)$, i.e., suppose $f_p$ is a feasible periodically repeated solution with $\mathcal{T}(f_p)=D/R_p$ and $\Lambda_p(f_p)=\Lambda_{p}^{R_p}$. Recall that follow the proof to Lem.~\ref{lem:MLP}, from $f_p$, we can construct a feasible solution $\textbf{f}(R_p)$ to $\textsf{MMD1}(R_p)$. For this constructed $\textbf{f}(R_p)$, an important observation of $\hat{\mathcal{M}}(\textbf{f}(R_p))$ is that $\mathcal{M}(f_p)\ge \hat{\mathcal{M}}(\textbf{f}(R_p))$, since that during the construction of $\textbf{f}(R_p)$ from $f_p$, (i) $\textbf{f}(R_p)$ does not use any path that is assigned zero data in $f_p$, and (ii) $\textbf{f}(R_p)$ decrease all the data-holding delays on nodes to zero based on $f_p$. Therefore, we have the following
	\bee
	\mathcal{M}(f_p) && \ge \hat{\mathcal{M}}(\textbf{f}(R_p))\overset{(a)}{\ge}\hat{\mathcal{M}}(\textbf{f}^*(R_p))\overset{(b)}{\ge}\frac{\hat{\mathcal{M}}(\textbf{f}_{\alpha}(R_{l}))}{\alpha}\nnb\\
	&& \overset{(c)}{=}\frac{\mathcal{M}(\textbf{f}_{\alpha}(R_{l}))-D/R_{l}+1}{\alpha},\label{eqn:p-m-e}
	\eee
	where inequality (a) comes from the optimality of $\textbf{f}^*(R_p)$ as to $\textsf{MMD1}(R_p)$, the inequality (b) holds due to the inequality~\eqref{eqn:epsilon}, and the equality (c) is true because of our Lem.~\ref{lem:MLP}.
	
	Now based on our Lem.~\ref{lem:age-delay} and the inequality~\eqref{eqn:p-m-e}, we have
	\bee
	\frac{\Lambda_p(\textbf{f}_{\alpha}(R_{l}))}{\Lambda_{p}(f_p)} && =\frac{\mathcal{M}(\textbf{f}_{\alpha}(R_{l}))+D/R_{l}-1}{\mathcal{M}(f_p)+D/R_p-1}\nnb\\
	&& \le\frac{\alpha\cdot\mathcal{M}(f_p)+2D/R_{l}-2}{\mathcal{M}(f_p)+D/R_p-1}\nnb\\ && =\frac{\alpha\cdot\mathcal{M}(f_p)}{\mathcal{M}(f_p)+D/R_p-1}+\frac{2(D/R_{l}-1)}{\mathcal{M}(f_p)+D/R_p-1}\nnb\\
	&& \overset{(a)}{\le} \alpha+\frac{2(D/R_{l}-1)}{D/R_p}\le \alpha+2\cdot\frac{R_{u}}{R_{l}},\nnb
	\eee
	where the inequality (a) comes from the fact that $\mathcal{M}(f_p)\ge 1$ and $D/R_p\ge 1$ since $d_e\in\mathbb{Z}^+$ for each link $e\in E$.
	Overall, we prove the approximation ratio $(\alpha+\mathsf{c})$.
	
	\textbf{Finally}, following the same method, we can prove the approximation ratio $(\alpha+\mathsf{c})$ of $\textbf{f}_{\alpha}(R_l)$ with $\textsf{MAA}(R_l,R_u,D)$.
\end{proof} 

\end{document}